\documentclass{article}
\usepackage[utf8]{inputenc}
\usepackage[english]{babel}
\usepackage{enumerate}
\usepackage{setspace}
\usepackage[margin=1.2in]{geometry}
\usepackage{authblk}
\usepackage{graphicx}
\usepackage{palatino}
\usepackage{xcolor}
\usepackage{amsmath}
\usepackage{amssymb}
\usepackage{float}
\usepackage{pdflscape}
\usepackage{afterpage}
\usepackage{tablefootnote}
\usepackage[titletoc,title]{appendix}
\usepackage[backend=bibtex,style=authoryear,maxcitenames=5, mincitenames=1]{biblatex}
\usepackage{etoolbox}
\makeatletter
\patchcmd{\Ginclude@eps}{"#1"}{#1}{}{}
\makeatother
\newcommand*\samethanks[1][\value{footnote}]{\footnotemark[#1]}
\usepackage{caption}
\newcommand\fnote[1]{\captionsetup{font=footnotesize}\caption*{#1}}

\title{Racial and Ethnic Disparities in Mortgage Lending: New Evidence from Expanded HMDA Data}
\author{Sean Lewis-Faupel\footnote{Office of the Comptroller of the Currency. The views expressed in this paper are those of the authors, do not necessarily reflect the views of the Office of the Comptroller of the Currency, the U.S. Department of the Treasury, or any federal agency, and do not establish supervision policy, requirements or expectations.} \and Nicholas Tenev\samethanks}
\date{January 19, 2024}

\begin{document}
	
	\maketitle
	
	\begin{abstract}
		This paper investigates gaps in access to and the cost of housing credit by race and ethnicity using the near universe of U.S. mortgage applications. Our data contain borrower creditworthiness variables that have historically been absent from industry-wide application data and that are likely to affect application approval and loan pricing. We find large unconditional disparities in approval and pricing between racial and ethnic groups. After conditioning on key elements of observable borrower creditworthiness, these disparities are smaller but remain economically meaningful. Sensitivity analysis indicates that omitted factors as predictive of approval/pricing and race/ethnicity as credit score can explain some of the pricing disparities but cannot explain the approval disparities.  Taken together, our results suggest that credit score, income, and down payment requirements significantly contribute to disparities in mortgage access and affordability but that other systemic barriers are also responsible for a large share of disparate outcomes in the mortgage market.
	\end{abstract}
	
	\thispagestyle{empty}
	\newpage 
	\setcounter{page}{1}
	
	\section{Introduction}
	For most, access to an affordable mortgage is necessary for homeownership (\cite{barakova2003does}). As a result, inequities in access to housing credit and the cost of that credit can lead to disparities in who owns a home. While disparities in homeownership may be undesirable in their own right, they can also have broader societal consequences. For instance, differences in homeownership rates can explain a large portion of the racial wealth gap (\cite{akbar2019racial}) and contribute to lower intergenerational wealth mobility among minorities (\cite{toney2021intergenerational}).\footnote{See also \textcite{collins2011race} for a historical account of racial differences in homeownership.} 
	
	In this paper, we catalog the state of racial and ethnic differences in mortgage application outcomes using individual application data that many mortgage lenders are required to report to U.S. regulators. In addition to application decisions, loan costs, and applicant-reported race and ethnicity, we observe creditworthiness variables including credit score, debt-to-income ratio, and loan-to-value ratio. These creditworthiness variables were only recently added to the mandatory data reporting and are missing from many studies of disparities in mortgage lending. We focus on data from 2018 and 2019, the first two years of expanded reporting.
	
	In examining these disparities, we do not aim to test for discrimination or illegal credit practices. Instead, our results more broadly describe disparities in mortgage access and prices. These gaps include the impact of any discrimination, but they may also result from other structural inequalities related to lending. For instance, sorting into loan products, varied propensity to solicit prices from multiple lenders, and the availability of comparable properties during an appraisal do not necessarily constitute discrimination but can contribute to differences in credit outcomes. In addition, we do not observe all creditworthiness factors that lenders consider, meaning that the disparities we highlight may be due at least in part to differences in variables not included in our analysis.
	
	Though we do not speak directly to the role of discrimination, our work contributes to the literature on discrimination in mortgage lending by suggesting a current upper bound on its market-wide impact.\footnote{To be clear, our results do not speak to an upper bound on discrimination within subpopulations (e.g., borrowers using a certain lender).} The existing literature on discrimination in mortgage lending is contentious. \textcite{munnell1996mortgage} augment Home Mortgage Disclosure Act (HMDA) data in Boston with voluntarily disclosed application information from lenders. The authors find large unconditional disparities in loan approval and smaller but economically meaningful disparities after conditioning on creditworthiness, a result similar   to ours despite being decades earlier and based on a limited sample.
	
	\textcite{munnell1996mortgage} spurred a literature debating their results, with many attempting to disprove the finding of disparities. For instance, \textcite{harrison1998mortgage} claims that the racial disparity disappears when other variables are included. \textcite{tootell1996redlining} finds that lenders are less likely to lend to minority applicants, but no evidence that the racial composition of the applicant's neighborhood matters. More recently, correspondence studies have become popular as a way to study discrimination. \textcite{hanson2016discrimination} conduct a correspondence experiment and find that mortgage lenders are less likely to respond to a request for information from a name that signals a Black applicant.

	\textcite{10.1257/jep.12.2.41} discusses and interprets prior results, distinguishing clearly between so-called taste-based discrimination and statistical discrimination. In our context, statistical discrimination entails lenders preferring non-Black borrowers because of creditworthiness alone. There may be context to support this hypothesis. For example, racial discrimination in job search or racial differences in wealth may affect the ability to repay of even those with spotless credit.\footnote{\textcite{bertrand2004emily} find racial differences in call-backs for job inquiries; \textcite{tenev2018social} finds racial differences in jobs found through friends; \textcite{derenoncourt2021racial} studies long-term racial differences in wealth.} In other words, headwinds in other areas of life may make statistical discrimination based on race profitable for lenders (though we stress that profitability is neither a legal nor a moral justification). \textcite{discrimdc} find that some loan officers do report believing there are racial differences in creditworthiness. However, \textcite{kau2012racial} find that conventional fixed-rate mortgage borrowers in predominantly Black neighborhoods pay higher rates than implied by their loan performance, suggesting that statistical discrimination alone cannot explain racial disparities in mortgage lending. 
	
	This paper also contributes to efforts to tabulate the newest data elements collected under HMDA. Extensive work by the Consumer Financial Protection Bureau summarizes trends and correlations in this data (\cite{cfpb2019}; \cite{cfpb2020}). We extend this work with a greater focus on race and ethnicity. \textcite{bartlett2021consumer} find that minority applicants pay more for government-guaranteed loans (FHA and GSE) that should not incur lender risk, though \textcite{bhutta2021pricing} argue using data on FHA loans that apparent pricing disparities may owe to differences in the propensity of different groups to pay up-front ``discount points’’ in exchange for lower interest payments. \textcite{popick2022did} finds pricing disparities using a simultaneous equation model incorporating both interest rates and discount points.  \textcite{zhang2021lenders} argue that racial and ethnic pricing disparities cannot be fully explained by differences in borrower choices, and owe at least in part to the ``menus’’ (combinations of points paid and subsequent interest rate) borrowers are offered. Compared to this recent work, our pricing analysis documents the extent to which aggregate disparities can be accounted for by differences in common pricing factors but leave the debate over points paid to other work.
	
	Our paper also documents disparities in denials, and a few projects developed concurrently with ours also use HMDA data and produce results similar to ours. \textcite{kylim2022} focus on disparities in denial in a subset of loans similar to our ``standard purchase’’ loan subsample (see Section \ref{sec:standardhp}; we also consider a wider sample). Though they exclude applications likely to be denied for low credit score, high loan-to-value (LTV), or debt-to-income (DTI),\footnote{Specifically, they exclude applications with credit score less than 620, LTV above 97\%, or back-end DTI above 50\%.} they find racial and ethnic disparities in denials that are often within a percentage point of ours. \textcite{bhutta2022bias}, focusing on applications that go through Automated Underwriting Systems (AUS), also find 1-2 percentage points of ``excess denials’’ for certain groups, though they argue that this may be accounted for by unobserved factors such as incomplete applications. \textcite{popick2022did} finds similar disparities in denials, and even larger disparities for a subsample of FHA loans. 
	
	Compared to the rest of the literature, our paper has a broader scope, including all mortgages other than HRB (home equity, reverse mortgages, or commercial or business-purpose loans) as well as seven distinct racial and ethnic categories, including American Indian or Alaska Native as well as Native Hawaiian or Pacific Islander---see column 4 in Table \ref{tab:comparison} for the number of groups considered by each paper. Another notable contribution relative to these papers (including those on pricing) is our sensitivity analysis (Section \ref{sec:sensitivity}), which gives a sense of how sensitive our results are to the potential presence of omitted variables correlated with race or ethnicity.  

	\begin{table}[h]
		\caption{Comparison of scope of work using HMDA 2018+ data \label{tab:comparison}}
		\begin{tabular}{lllll}
			& Years  & Outcomes & \# Groups & Subsample(s) \\
			\hline 
			\textcite{bartlett2021consumer}$^\dagger$ & 2009-15; & Pricing & 2 & GSE; FHA \\
			& 2018-19 & & &\\
			\textcite{bhutta2021pricing}$^\dagger$ & 2018-19  & Pricing  & 5 & FHA \\
			\textcite{zhang2021lenders} & 2018-19 & Pricing & 3 &  30y conf. HP; FHA\tablefootnote{More precisely, \textcite{zhang2021lenders} restrict their analysis to 30-year first-lien fixed-rate purchase loans for owner-occupied, site-built properties with no prepayment penalties, balloon, interest-only, negative amortization, or non-amortizing features.}\\
			\textcite{bhutta2022bias} & 2018-19 & Denial & 5 & Fixed-rate 30y AUS \\
			\textcite{kylim2022}& 2018-20  & Denial & 4 & 30y conv. HP\tablefootnote{More precisely, \textcite{kylim2022} focus on 30-year conforming first-lien mortgages for purchase of single-dwelling units for primary residence.}  \\
			\textcite{popick2022did} & 2020 & Denial, & 3 & Conv. conf. HP;  \\
			& & Pricing & &FHA HP; conf. refi.\tablefootnote{\textcite{popick2022did} further splits conforming refinance subsamples into cash-out and no cash-out.}  \\
			Lewis-Faupel and Tenev (2022) (this paper) & 2018-19 & Denial, & 7 & non-HRB;  \\
			& & Pricing & &30y conv. HP\tablefootnote{For the specifics of our ``standard purchase'' subsample, see \ref{sec:standardhp}} \\
			\hline
		\end{tabular} \\
		
		\fnote{\footnotesize{$^\dagger$These papers do not use the full confidential HMDA data, and use e.g. credit score data from other data sets. ``\# Groups'' denotes the number of distinct racial and ethnic groups considered in the analysis.}}
		
	\end{table}
	
	Section \ref{sec:data} describes the data used in this paper, and Section \ref{sec:disparitygraphs} demonstrates graphically that racial and ethnic disparities exist both in aggregate outcomes as well as for borrowers with similar characteristics. Section \ref{sec:regressions} analyzes these disparities through the lens of linear regressions, which demonstrate the extent to which variation in outcomes can be accounted for by variation in common underwriting and pricing variables. However, Section \ref{sec:sensitivity} shows that some of these results may be sensitive to variables that impact underwriting or pricing and are correlated with race or ethnicity but not included in our regressions. Section \ref{sec:discussion} discusses the results, and the Appendix provides ancillary material. 
	
	\section{Data}\label{sec:data}
	Our analysis relies on data reported by financial institutions under the Home Mortgage Disclosure Act or HMDA. Since its passage in 1975, HMDA has required certain lenders to report requests for credit related to residential real estate.\footnote{See \textcite{HMDAcoverage} for a complete description of reportable transactions during our study period.} The set of institutions required to report this data and the scope of the data reported has expanded through a series of amendments to the Act. During our study period, banks, credit unions, savings associations, and for-profit mortgage lending institutions must report data if they have at least 25 closed-end mortgage originations or 500 open-end lines of credit in the previous two years. Lenders without an office or branch presence in an MSA or with no mortgage loans in an MSA are exempt.\footnote{See \textcite{HMDAinstitutions} for all coverage criteria. Among lenders required to report HMDA data, those with fewer than 500 closed-end mortgage originations in each of the two preceding years and that meet certain other criteria, may opt to not report certain datapoints, including some of those used in our analysis.}  Analysis by the Consumer Financial Protection Bureau suggests that around half of mortgage lenders report to HMDA and that these lenders constitute around 90 percent of mortgage lending in the U.S. (\textcite{datapoint2019}).
	
	Beginning in 1989, HMDA and its implementing regulations required lenders to report application-level data. In addition to reporting the outcome of each application, lenders are required to ask applicants about their race and ethnicity and report that information for up to two applicants for each application. Starting in 2018, requirements expanded to include reporting several creditworthiness factors if relied on to make a credit decision. These new additions include applicant's credit score, debt-to-income ratio (DTI), and loan-to-value ratio (LTV).
	
	\subsection{Race and ethnicity in HMDA}\label{sec:raceethHMDA}
	
	\afterpage{
		\begin{landscape}
			\begin{figure}[ht]
				\centering
				\includegraphics[scale=0.09]{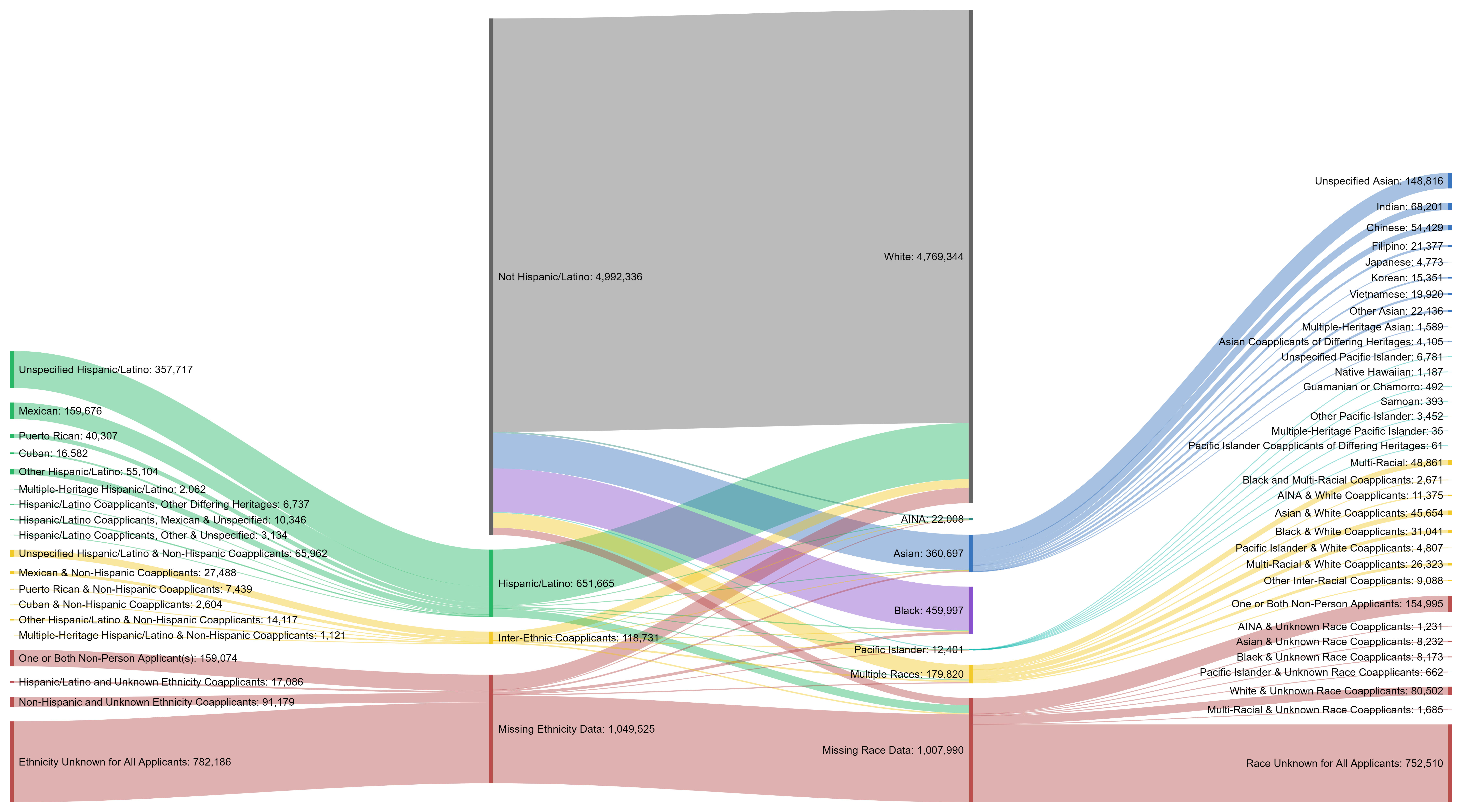}
				\caption{Reported Race and Ethnicity of Applicants}
				\label{fig:raceandeth}
			\end{figure}
		\end{landscape}
	}
	
	HMDA requires that lenders ask for information about borrowers' race. Specifically, ``Regulation C, 12 CFR $\S$ 1003.4(a)(10)(i), requires that a financial institution collect the ethnicity, race, and sex of a natural person applicant or borrower, and collect whether this information was collected on the basis of visual observation or surname'' (\textcite{hmdafaq}). \textcite{office1997revisions} describes the principles used in developing the categorization, and summarizes some of the discussion and dissent surrounding the choices made. The race and ethnicity categories recorded in HMDA are listed in Section \ref{sec:HMDArace}. From 2018 onwards, the HMDA data allow, for the first time, applicants who are Hispanic or Latino, Asian, or Native Hawaiian or Pacific Islander to report finer ethnicity categories within these groups. Amongst our sample, 35\% of applications reporting applicant or coapplicant race as Asian did not report a finer categorization, 46\% of Native Hawaiian or Pacific Islander reports did not specify a finer categorization, and 52\% of those reporting Hispanic or Latino ethnicity did not include a finer categorization.
	
	Mortgage applications can list a co-applicant as well as an applicant, and each can report multiple racial and ethnic categories (or write in their own). To present results for all possible combinations would be infeasible, so in this paper we categorize an application as belonging to a certain race and ethnicity only if no other race or ethnicity was reported on the application.\footnote{For applicants that report a certain race as well as a subcategory of that race, we classify them as the subcategory (e.g., we classify as Korean an applicant that ticks Asian as well as Korean).} We also report a category for those applications with multiple race or ethnic categories listed, and another for applications missing all race or ethnicity data.
	
	It is important to note that race and ethnicity are distinct categorizations in HMDA, elicited by separate questions. Figure \ref{fig:raceandeth} shows the breakdown of ethnicity on the left, and the breakdown of race on the right, with broader categories in the middle and finer categorizations at the two sides.
	
	The vast majority of applicants who report Hispanic/Latino ethnicity report their race as White. Accordingly, we separate applicants who report their race as White by ethnicity into Non-Hispanic/Latino White and Hispanic/Latino. The fraction of HMDA applicants with no reported ethnicity largely overlaps with those with no reported race. Accordingly, in addition to categories for each race group (with White broken down by Hispanic/Latino ethnicity), we report a ``Multiple Responses'' category as well as an ``All Missing'' category.
	
	For a full accounting of how we categorize race and ethnicity, see Appendix Section \ref{sec:racecat}.
	
	\subsection{Standard purchase loan applications}\label{sec:standardhp}
	
	We define ``standard’’ purchase loans,  sometimes abbreviated as ``StdHP,'' as 30-year conventional\footnote{HMDA reporting guidelines define ``conventional'' as not insured or guaranteed by FHA, VA, RHS, or FSA.} fully-amortizing home purchase loans secured by first-lien single-family properties, excluding home equity loans, reverse mortgages, and loans for commercial or business purposes.\footnote{Adding an indicator for loan amounts above the national conforming loan limit in our controls does not significantly change the results; the same is true for the pricing results in Table \ref{tab:ratereg}.\label{fn:conform}} To control for differences between loan products, in the following results we will often restrict attention to applications that meet these criteria, which constitute the most popular mortgage loan products. However, we will also report results for a broader set of applications, ``Non-HRB,’’ that excludes only home equity, reverse mortgages, and loans for commercial or business purposes.

\begin{figure}[H]
		\includegraphics[trim=2 2 2 2, clip, scale=1]{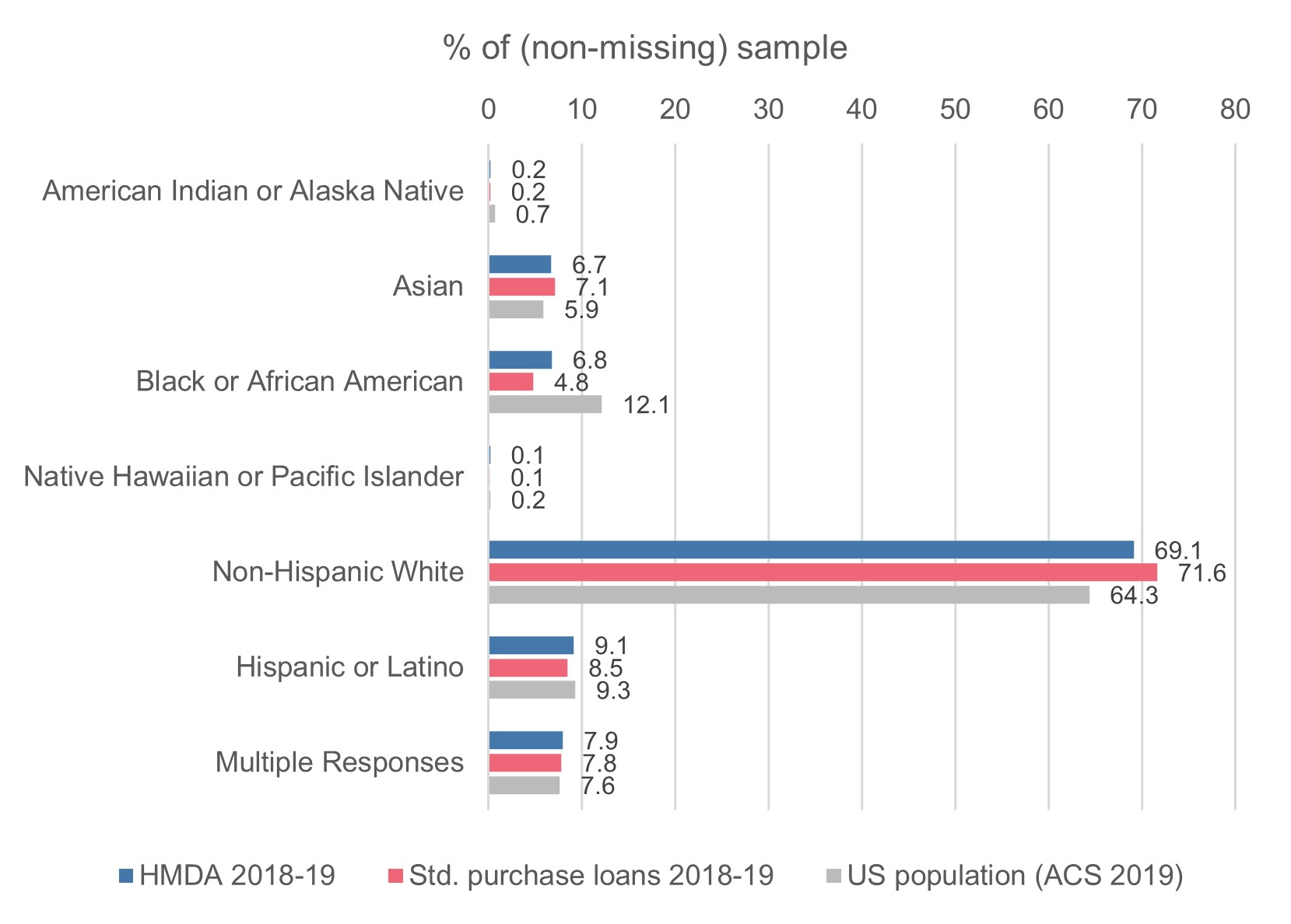}
		\caption{Mortgage application sample by race/ethnicity, compared to US population}
		
		\centering
		\fnote{Note: 11.9\% of the HMDA mortgage application sample and 9.6\% of the standard purchase loan sub-sample were missing race/ethnicity information; the percentages shown above are fractions of the non-missing sample. US population data include individuals 18 and older, and are from the American Community Survey.}
		\label{fig:appsrace}
	\end{figure}

	Figure \ref{fig:appsrace} shows the proportion of applications in our sample by race/ethnicity (as defined above), for both the full sample as well as the subset we define as standard purchase loans. The composition of the US population is provided as well, to show which groups are over- and under-represented in mortgage applications. 
	The majority group (Non-Hispanic White) is over-represented in the mortgage application data compared to their share of the population, as are Asian applicants, while the other groups are under-represented compared to their shares of the population. However, owing to the sizeable fraction of applications missing race/ethnicity data, it is hard to draw firm conclusions about the degree of representation of any group.
	
	\section{Visualizing disparities by race and ethnicity}\label{sec:disparitygraphs}
	
	\afterpage{
		\begin{figure}[H]
			\includegraphics[trim=2 2 2 2, clip, scale=1.0]{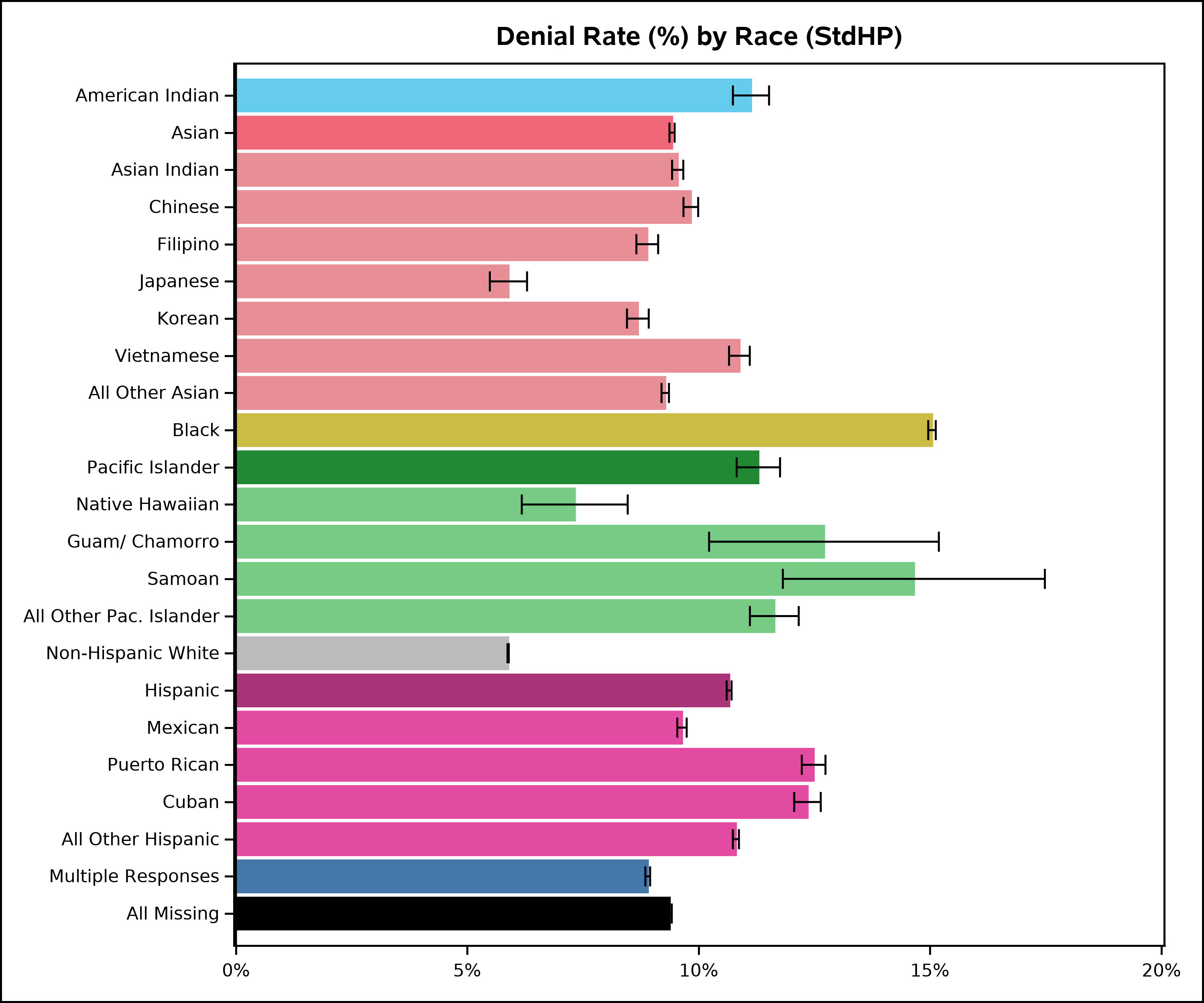}
			\centering
			\caption{Denial rate of standard home purchase applications by race/ethnicity}
			\label{fig:denialrace}
		\end{figure}
	}
	
	Figure \ref{fig:denialrace} shows the denial rate of standard home purchase applications by race and ethnicity (as defined above). Black, Samoan, Guamanian/Chamorro, Puerto Rican, and Cuban applicants have the highest denial rates among these groups, while non-Hispanic White and Japanese applicants have the lowest denials rates. While this figure includes all categories described above to demonstrate the heterogeneity in outcomes across subgroups, the remainder of our results will aggregate these subgroups to six racial and ethnic groups as described in the Section \ref{sec:racecat}, as well as Missing and Multiple Responses.
	
	These raw disparities in the chances of approval can be explained in part by group differences in the variables lenders use to approve mortgage loans. In particular, this paper will focus on three of the most common underwriting variables used by lenders: credit score, which in most cases is one of several standard scores provided to the lender by a major credit reporting agency; the combined loan-to-value ratio (CLTV), which is indicative of the size of the down payment; and the debt-to-income ratio (DTI), which is determined by income and total debt payments. 
	
	Even for applicants with similar credit scores, the CLTV and DTI of mortgage applications vary by race and ethnicity. Figure \ref{fig:CLTV_race_score} shows the average CLTV by race and credit score. 
	
		\afterpage{
		\begin{figure}[H]
			\textcolor{white}{
				\includegraphics[trim=2 2 2 2, clip, scale=.95, page=2]{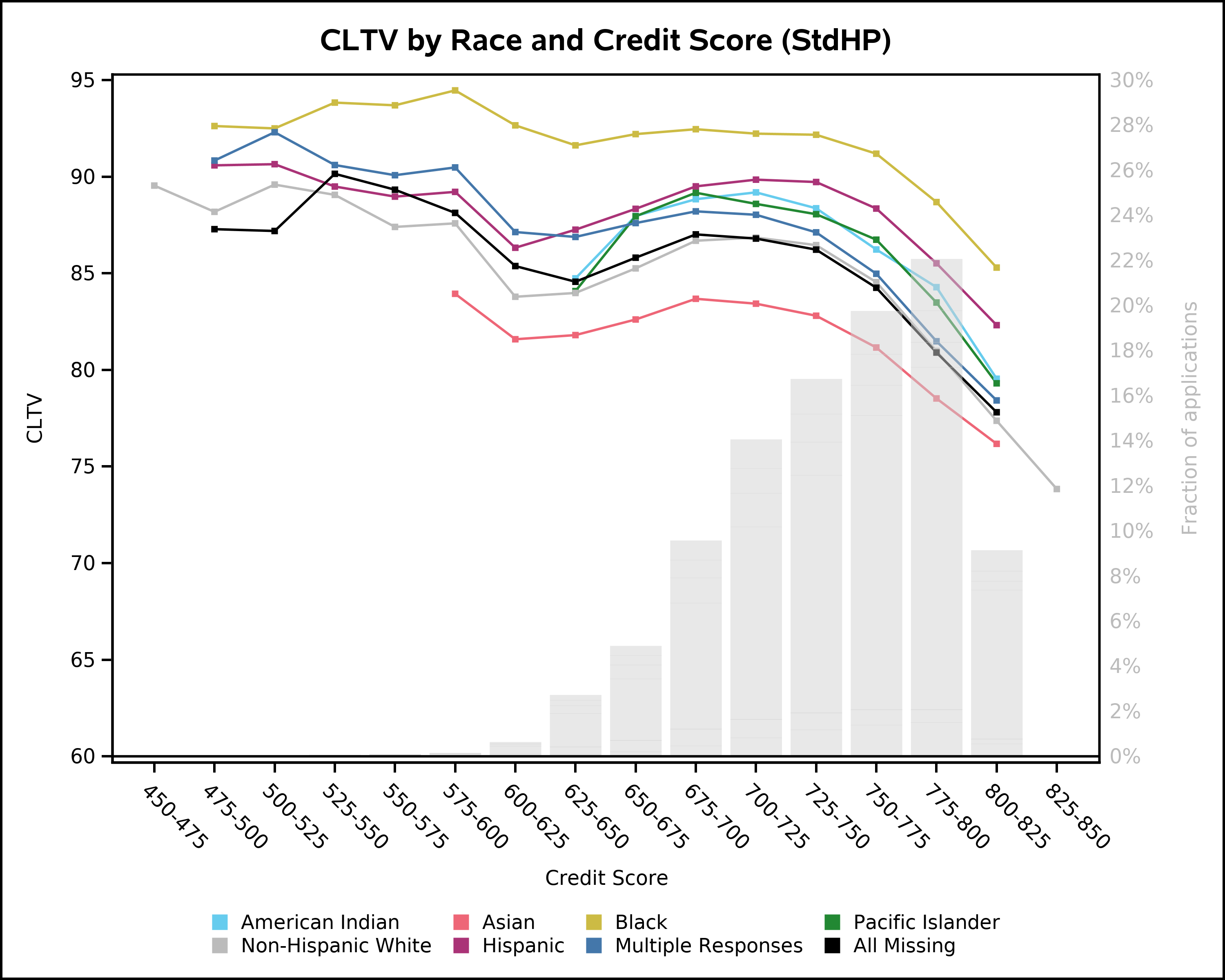}
			}
			\centering
			\caption{Average CLTV of standard home purchase applications by credit score and race/ethnicity}
			\label{fig:CLTV_race_score}
		\end{figure}
		
		\begin{figure}[H]
			\textcolor{white}{
				\includegraphics[trim=2 2 2 2, clip, scale=.95, page=4]{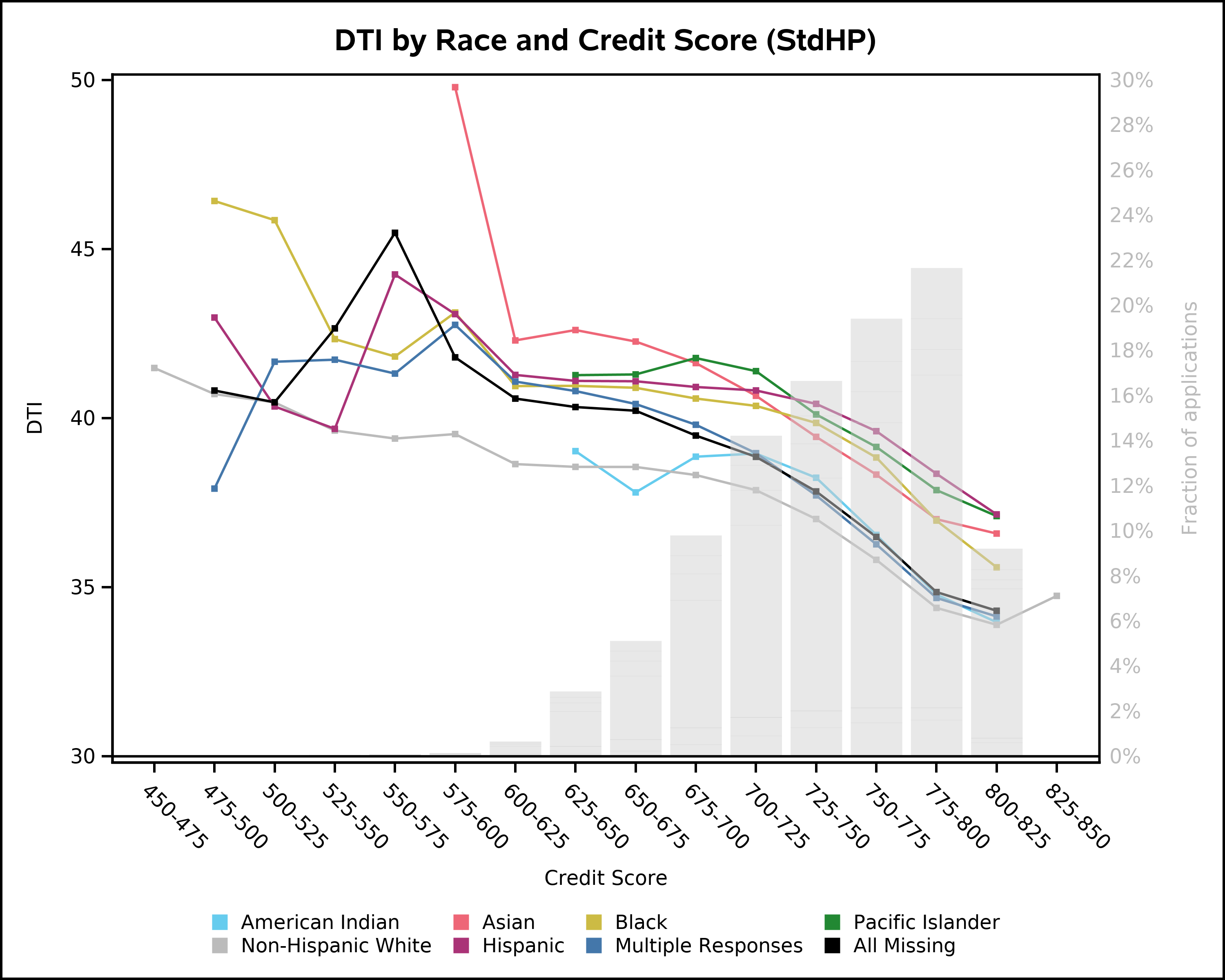}
			}
			\centering
			\caption{Average DTI of standard home purchase applications by credit score and race/ethnicity}
			\label{fig:DTI_race_score}
		\end{figure} 
	}
	
	In this and subsequent figures, the density of applications (by x-axis bin) is plotted in grey bars (scale on right axis), while the outcome of interest (in this case, average CLTV) is plotted with one line for each race or ethnic group (scale on left axis). The minimum sample size for a point to be plotted is 400. Hence the line for non-Hispanic White applicants, for example, extends further than the others because that group has sufficient sample size across the credit score distribution.
	
	While borrowers with higher credit scores tend to have lower CLTV, a noticeable racial and ethnic disparity in CLTV is consistent across the spectrum of credit scores. In particular, Black applicants have significantly higher CLTV--even those with high credit scores. This may reflect the racial wealth gap\footnote{\textcite{derenoncourt2021racial} measure the racial wealth gap going back to 1860.}; borrowers with less wealth may rely on financing for a greater fraction of the purchase price of a new home.

	Figure \ref{fig:DTI_race_score} shows the average debt-to-income (DTI) ratio by race and credit score. In general, applicants with higher credit scores tend to have lower DTI ratios. While racial and ethnic disparities exist for borrowers with similar credit scores, these disparities may not fully reflect the mortgage consequences of gaps in racial and ethnic income levels, since we also find that lower income individuals apply for small amounts of mortgage debt. 
	
	\subsection{Denial rates by application characteristics}
	This section shows how denial rates for standard home purchase applications vary by common creditworthiness characteristics: credit score, debt-to-income ratio, and loan-to-value ratio.

	\begin{figure}[H]
		\textcolor{white}{
			\includegraphics[trim=2 2 2 2, clip, scale=.95]{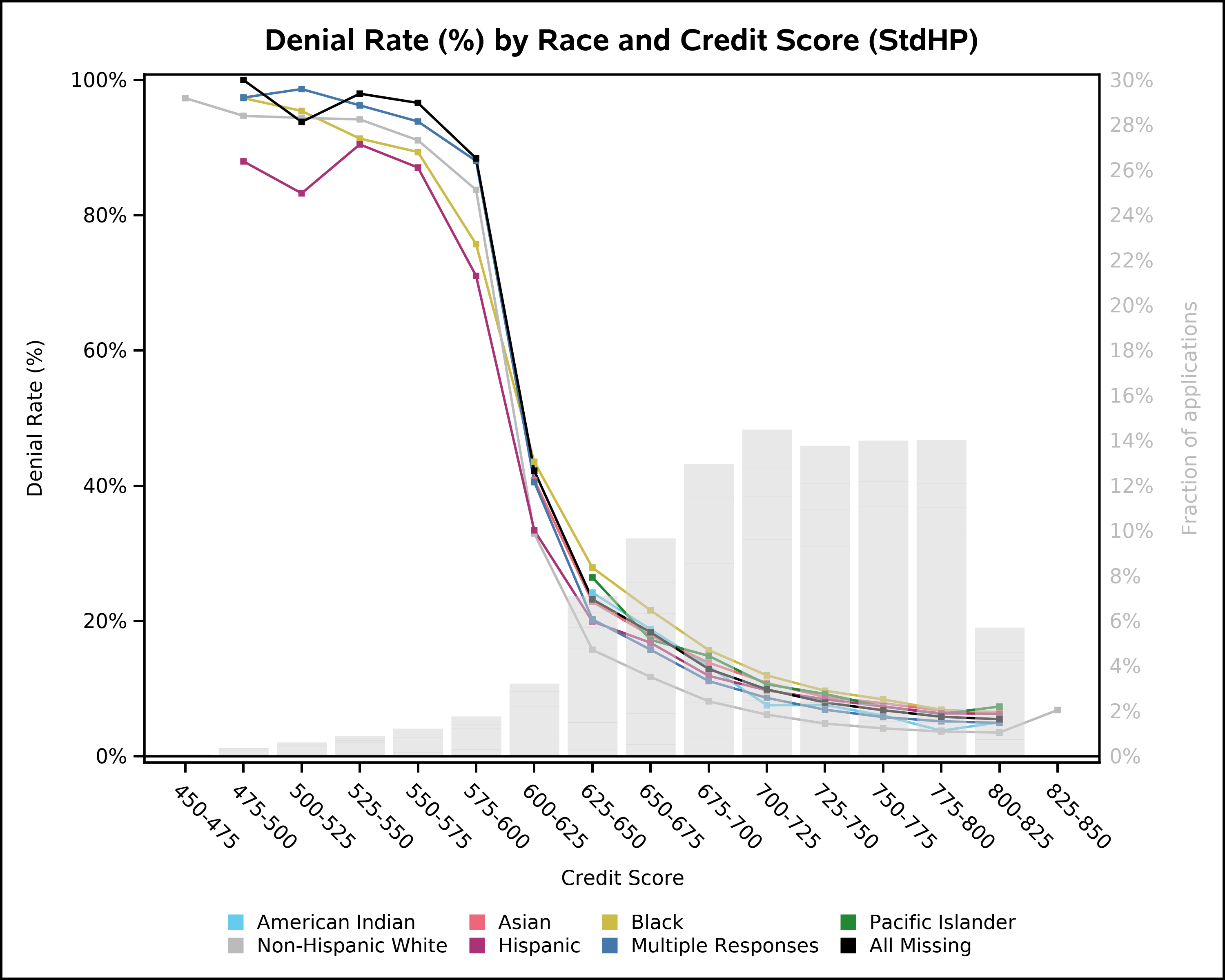}
		}
		\centering
		\caption{Denial rate of standard home purchase applications by credit score and race/ethnicity}
		\label{fig:denialracescore}
	\end{figure}
	
	The denial rate varies dramatically with credit score (Figure \ref{fig:denialracescore}). For all races and ethnicities, denial is much less likely for applicants with credit scores above 600, and the probability of denial generally continues to decline with credit score. This pattern is very likely driven by credit score requirements in the secondary mortgage market. For example, the two largest government-sponsored purchasers of mortgages required a 620 credit score during this period.	

		\begin{figure}[H]
			\includegraphics[trim=2 2 2 2, clip, scale=.95]{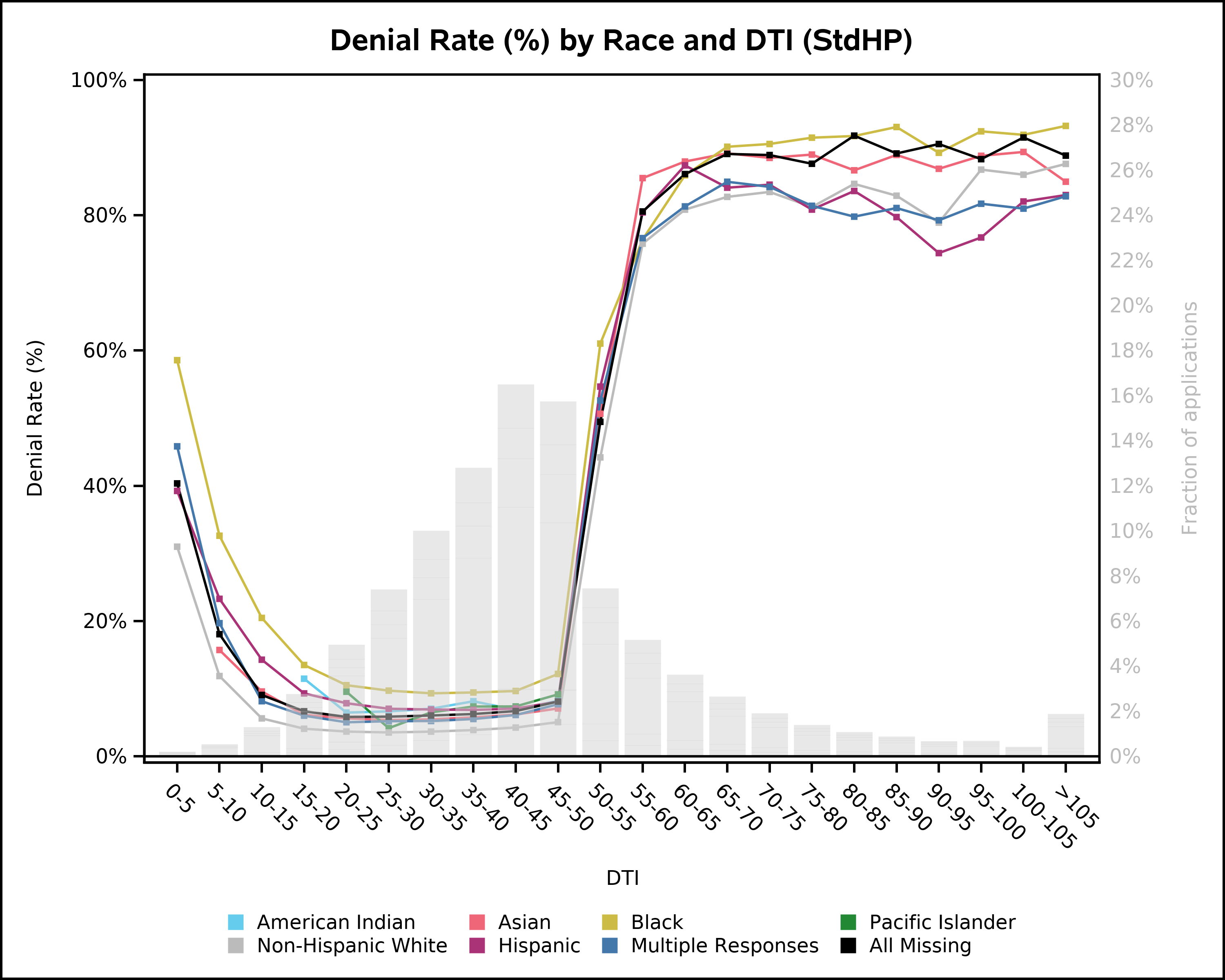}
			\centering
			\caption{Denial rate of standard home purchase applications by DTI and race/ethnicity}
			\label{fig:denialraceDTI}
		\end{figure}	
	For all race/ethnicity groups, denial is much more likely when the applicant's debt-to-income ratio exceeds 50\% (Figure \ref{fig:denialraceDTI}). Applicants may be aware of this and able to control it (by either not applying, or not seeking loans that would require debt payments of more than half their income)---there is noticeable bunching of applicants just below 50\% DTI, as seen in the grey bars (scale on right axis) which give the fraction of applicants at each level of DTI.	
	
\begin{figure}[H]
			\includegraphics[trim=2 2 2 2, clip, scale=.95]{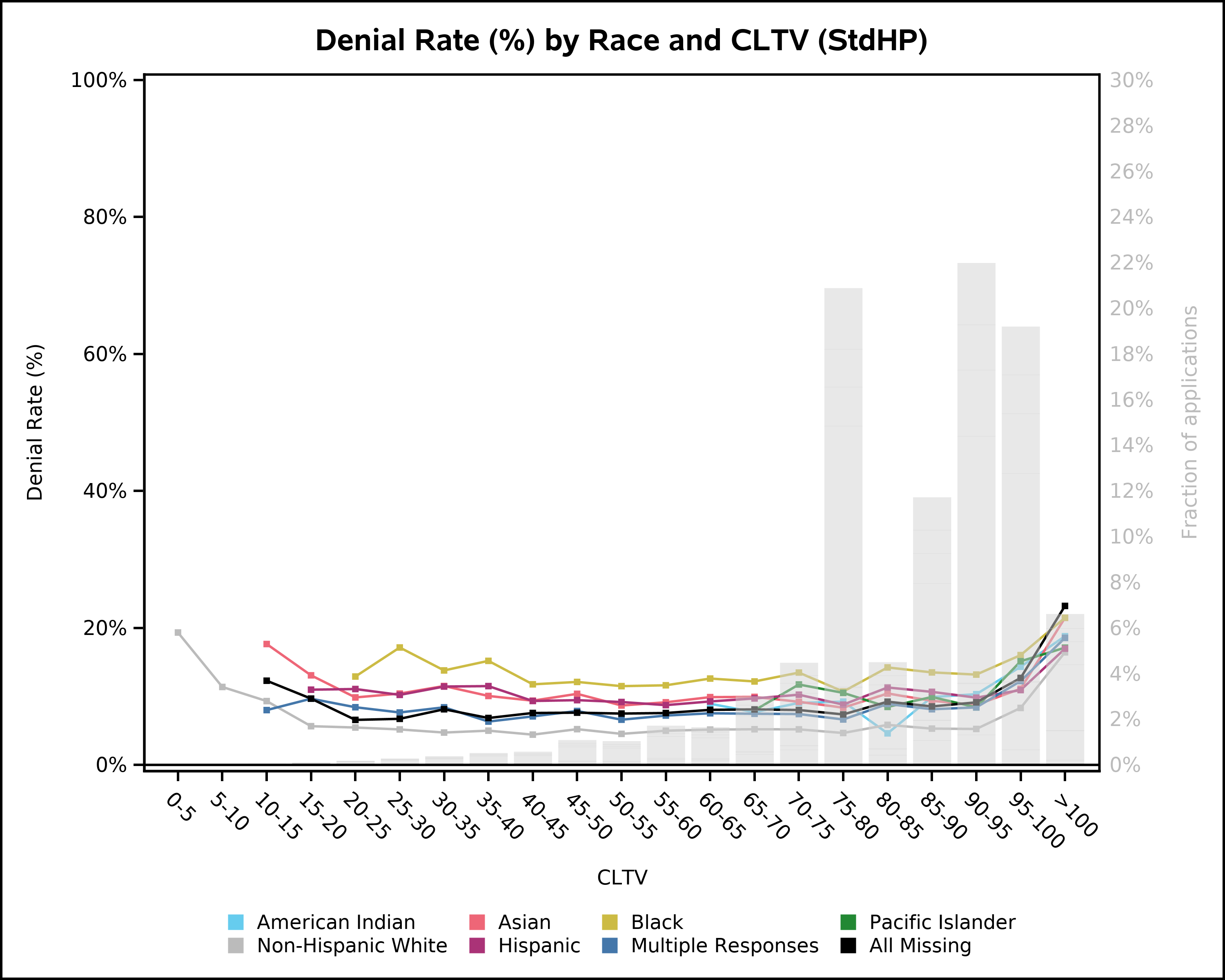}
			\centering
			\caption{Denial rate of standard home purchase applications by CLTV and race/ethnicity}
			\label{fig:denialraceCLTV}
		\end{figure}

	Compared to credit score and DTI, CLTV seems to have a much less dramatic effect on denial (Figure \ref{fig:denialraceCLTV}). Denial rates are higher for applicants seeking to borrow more than the value of their property, but the difference is much less than crossing the 50\% threshold in DTI, for example.
		
	The distribution of CLTV is bimodal: borrowers bunch just below the 80\% cutoff and, to a lesser extent, just below the 95\% cutoff. Despite the weak association between CLTV and approval, this suggests that before applying, applicants may be aware   of lender expectations and requirements for approval and are able to shape their applications to meet them (by, for example, applying for no less than a 20\% down payment). Applicants may also tend toward default values offered by lenders. For example, even borrowers for whom a 25\% down payment is optimal may end up with a 20\% down payment when 20\% is the default. 
	
	\subsection{Interest rate by application characteristics}

\begin{figure}[H]
		\textcolor{white}{
			\includegraphics[trim=2 2 2 2, clip, scale=.95]{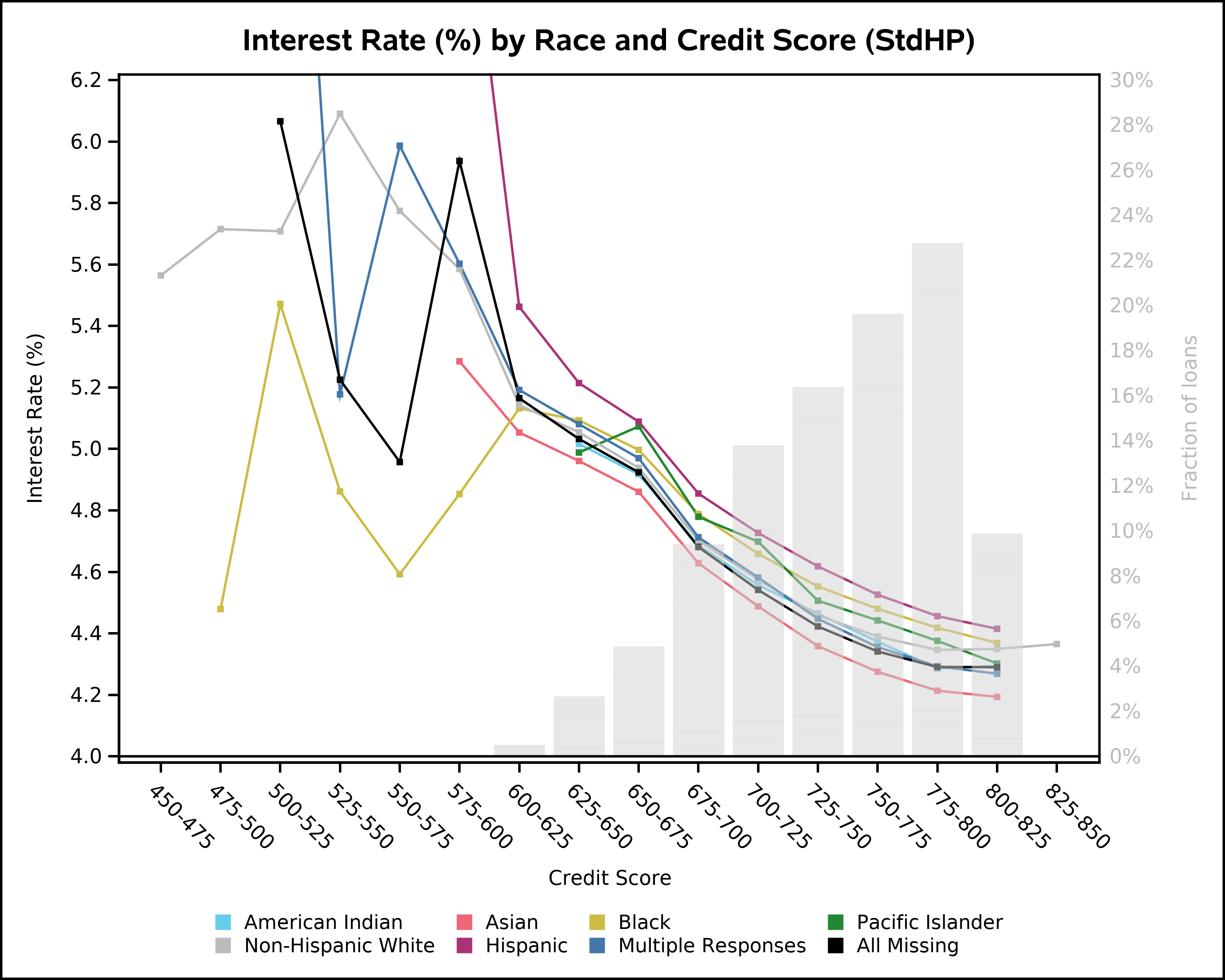}
		}
		\centering
		\caption{Interest rate of standard home purchase loans by credit score and race/ethnicity}
		\label{fig:rate_race_score}
	\end{figure}

	This section shows how interest rates for originated standard home purchase loans vary by common creditworthiness characteristics: credit score, debt-to-income ratio, and loan-to-value ratio.	Figure \ref{fig:rate_race_score} shows, for each race and ethnicity, how the interest rate varies with the applicant's credit score. 
	
	As expected, borrowers with higher credit scores pay lower interest rates. Credit score is unable to fully account for racial and ethnic disparities in interest rates, however. At any given credit score, some racial and ethnic groups pay on average higher rates of interest than others. This may owe in part to racial and ethnic differences in cash  paid at closing in exchange for a lower interest rate (known as discount points); \textcite{zhang2021lenders} examine this issue further.	


		\begin{figure}[H]
			\includegraphics[trim=2 2 2 2, clip, scale=.95]{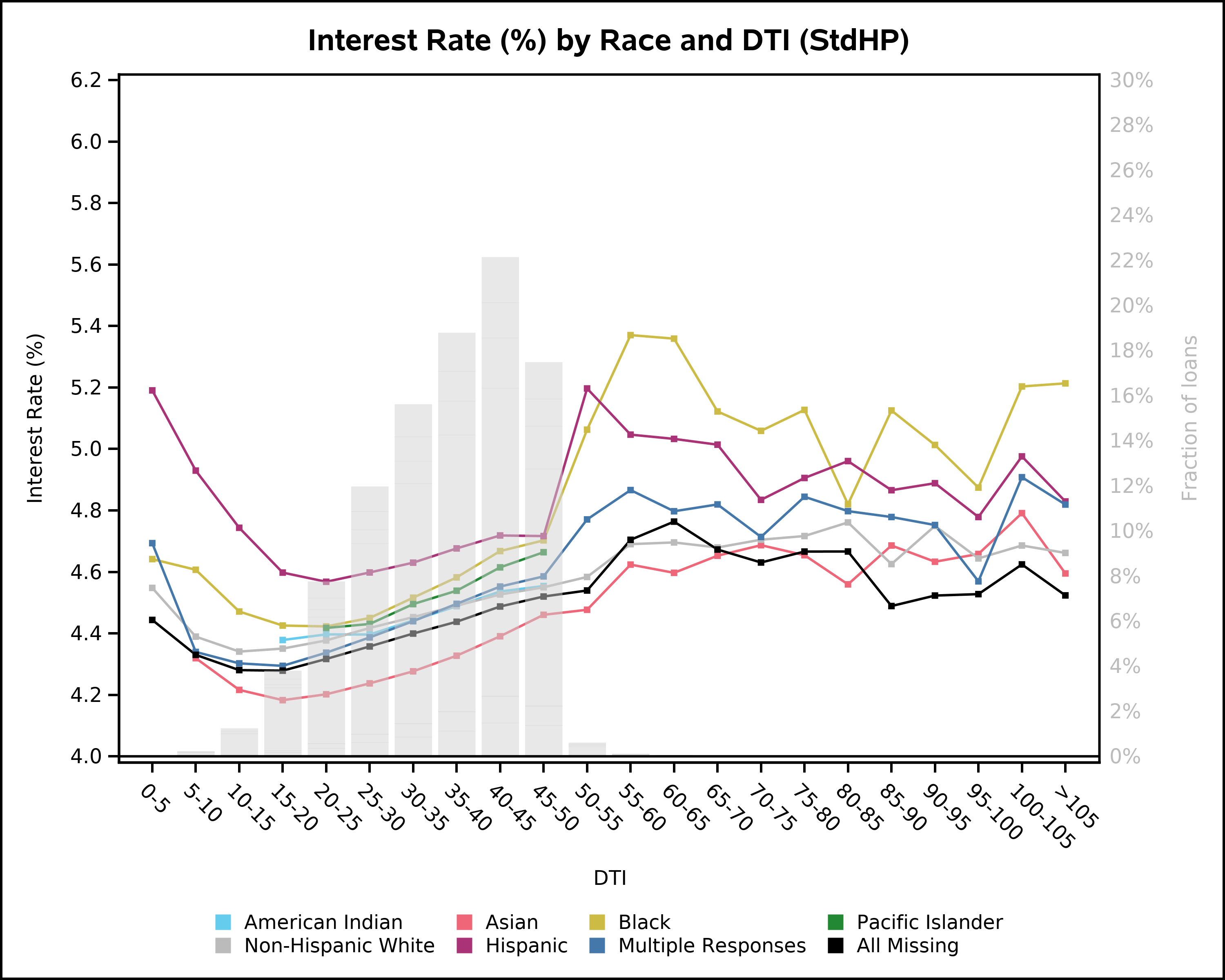}
			\centering
			\caption{Interest rate of standard home purchase loans by DTI and race/ethnicity}
			\label{fig:rateraceDTI}
		\end{figure}	
	Figure \ref{fig:rateraceDTI} plots average interest rate by DTI for each race and ethnicity. For most races and ethnicities, the interest rate does not vary much with DTI. For Black borrowers and to a lesser extent Hispanic or Latino borrowers, there does seem to be an increase in interest rate charged just above the 50\%, though few applicants are approved with such a high DTI (see Figure \ref{fig:denialraceDTI}). 		
		\begin{figure}[H]
			\includegraphics[trim=2 2 2 2, clip, scale=.95]{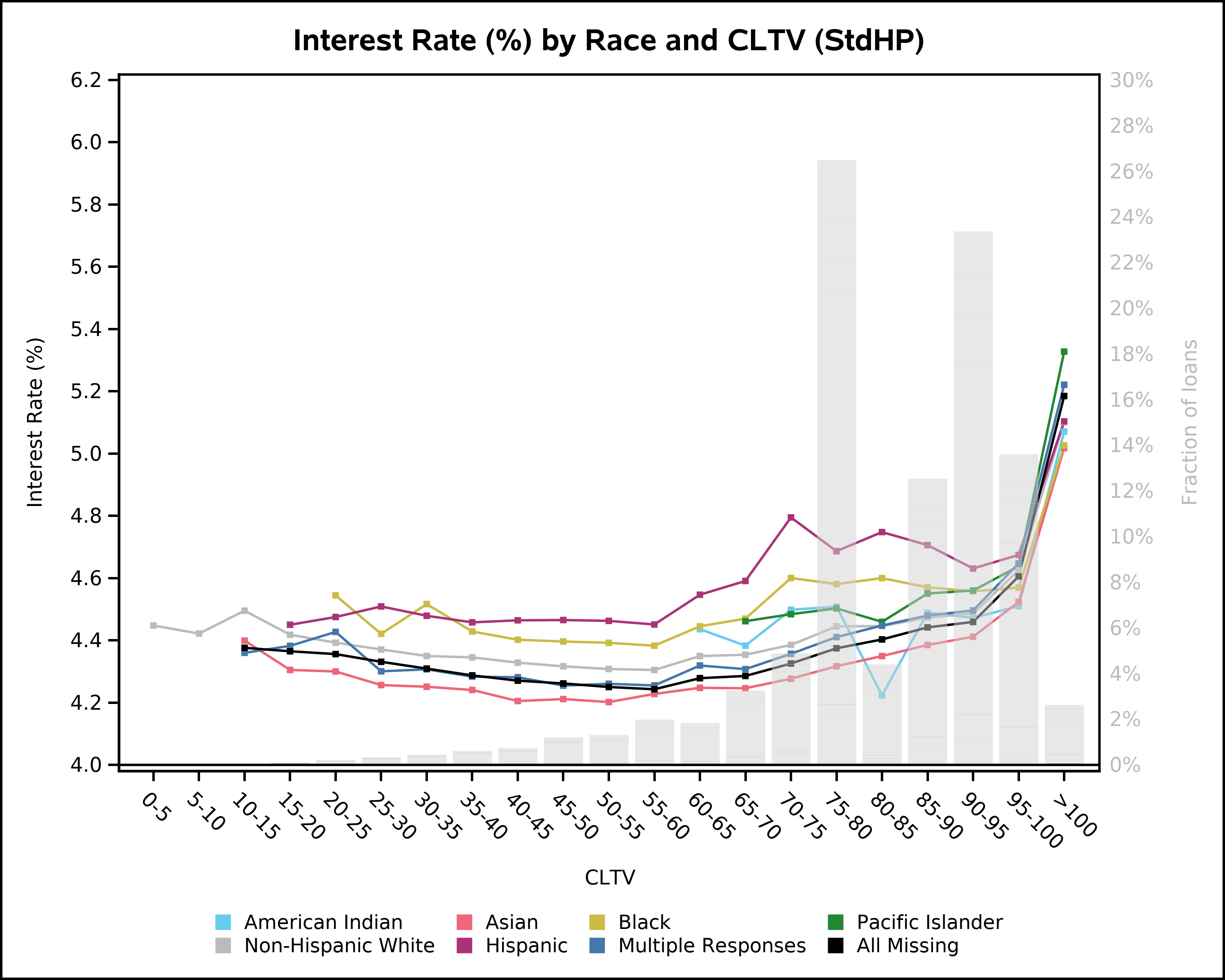}
			\centering
			\caption{Interest rate of standard home purchase loans by CLTV and race/ethnicity}
			\label{fig:rateraceCLTV}
		\end{figure}
	Similar to DTI, the CLTV of an application (Figure \ref{fig:rateraceCLTV}) seems to have only a modest association with the interest rate of the loan, except for high CLTVs above 95\%. 
	
	\section{Regression analysis}\label{sec:regressions}
	
	In the previous section, we examined denial probability and interest rate disparities visually along individual creditworthiness dimensions. To measure disparities simultaneously holding multiple creditworthiness factors constant, we turn to linear regression.\footnote{Parametric regressions for binary outcomes, such as logit and probit, are computationally infeasible with the number of parameters we estimate for some of our specifications. Furthermore, they are known to suffer from bias when the subsample that identifies a parameter is particularly small, which we encounter in several of our more flexible models.} We run OLS regressions of the following general form:
	
	$$ y_i = \beta' PBG_i + \gamma' X_i + \varepsilon_i $$
	
	where $y$ is either interest rate or an indicator for denial, $PBG$ (``prohibited basis group'') is a vector of mutually exclusive race/ethnicity indicators, and $X$ is a vector of controls.

	We consider two samples: all applications excluding home equity, reverse mortgages, and business/commercial purpose loans (HRB), and a narrower sample restricted to standard home purchase applications as defined earlier. For each of these two samples, we run five regressions. The first includes no controls except race and ethnicity, to measure the raw disparities. Second, we include credit score, DTI, and CLTV as linear controls. The third regression includes these three variables as binned controls (10-point intervals for credit score and 5-point intervals for DTI and CLTV), and the fourth includes all interactions of the binned controls. The fifth regression includes the same controls as the fourth and adds lender fixed effects.\footnote{\textcite{bhutta2022bias} run similar regressions on denial outcomes but also include county, month, and automated underwriting system (AUS) result as controls. \textcite{kylim2022} ’s regressions are also similar, but include state and time fixed effects as well as log loan amount and log applicant income. \textcite{popick2022did} includes state-MSA indicators and month fixed effects, as well as indicators for AUS, broker loan, and the presence of other liens, and uses the maximum credit score if there are two applicants. For pricing, \textcite{popick2022did} also includes income to proxy for borrower wealth.} 
	
	The goal of this exercise is not to establish a causal relationship between race/ethnicity and the outcomes. Instead, the regression analysis allows us to say how much of the raw disparities in denial rates (Figure \ref{fig:denialrace}) and interest rates (Figure \ref{fig:rate_race_score}) can be explained by the three major underwriting factors. The remaining disparities are not necessarily indicative of discrimination, since we do not observe all creditworthiness factors or other legal underwriting and pricing inputs. Our main results are those for the Standard Purchase subsample---loans which are more likely to have been underwritten and priced similarly. Our results for the larger sample (excluding only HRB) demonstrate the extent to which the same limited set of controls can account for disparities in aggregate outcomes.  
	
	Even for the Standard Purchase sample, we do not include as exhaustive a set of controls as our data allows. Instead, we only control for the three factors described above and omit some factors that we observe and that influence underwriting and pricing in certain cases. We do so because those observed but omitted factors are less universally determinative of our outcomes than credit score, DTI, and CLTV. As a result, including them may shift the coefficients on minority race and ethnicity in a way that complicates interpretation. For example, some lenders adjust interest rates based on geography, while others use uniform nationwide pricing (and we do not observe which lenders use local versus nationwide pricing). In this context and under certain other conditions, controlling for geography will overestimate the average relationship between geography and interest rates and underestimate the minority race/ethnicity coefficients.\footnote{While we do not establish general conditions under which inclusion of an observed factor $w$, which affects the outcome when omitted indicator $z$ is equal to 1, would downward bias $\beta_r$, the coefficient on race/ethnicity $PBG_r$, we note that there are reasonable values of $cov(PBG_r,z)$, $cov(PBG_r,w)$, $\gamma_{w|z=1}$ (the effect of $w$ when $z=1$), and $\beta_r$ for which $E[\hat{\beta}_r] > \beta_r$ when $w$ is omitted from the regression and $E[\hat{\beta}_r] < \beta_r$ when $w$ is included.} This occurs because the geography term captures not only the relationship with interest rate from the portion of our data where lenders price on geography but also any relationship between geography, race, and lender. In other words, disparities that are more directly related to race/ethnicity would be mistaken for geographic pricing variation because borrowers sort into geographies and lenders by race and ethnicity and because we cannot accurately condition on lenders’ use of geographic pricing policies. Ultimately, an underestimate of the coefficients on minority races and ethnicities would prevent interpretations of those parameters as any type of upper bound.

\afterpage{
	\begin{landscape}
		\begin{table}[!h]
\caption{Probability of Denial (\%), Relative to Non-Hispanic White\label{tab:denialreg}}
\centering
\begin{tabular}{lllllllllll}
\cline{1-11}
\multicolumn{1}{c}{} &
  \multicolumn{5}{|c}{All Excluding HRB} &
  \multicolumn{5}{c}{Standard Purchase} \\
\multicolumn{1}{c}{} &
  \multicolumn{1}{|r}{(1)} &
  \multicolumn{1}{r}{(2)} &
  \multicolumn{1}{r}{(3)} &
  \multicolumn{1}{r}{(4)} &
  \multicolumn{1}{r}{(5)} &
  \multicolumn{1}{r}{(1)} &
  \multicolumn{1}{r}{(2)} &
  \multicolumn{1}{r}{(3)} &
  \multicolumn{1}{r}{(4)} &
  \multicolumn{1}{r}{(5)} \\
\cline{1-11}
\multicolumn{1}{l}{Am. Indian/Alaska Native} &
  \multicolumn{1}{|r}{8.7***} &
  \multicolumn{1}{r}{5.7***} &
  \multicolumn{1}{r}{4.9***} &
  \multicolumn{1}{r}{4.4***} &
  \multicolumn{1}{r}{3.3***} &
  \multicolumn{1}{r}{4.4***} &
  \multicolumn{1}{r}{2.9***} &
  \multicolumn{1}{r}{1.9***} &
  \multicolumn{1}{r}{1.9***} &
  \multicolumn{1}{r}{1.8***} \\
\multicolumn{1}{l}{} &
  \multicolumn{1}{|r}{(0.2)} &
  \multicolumn{1}{r}{(0.2)} &
  \multicolumn{1}{r}{(0.2)} &
  \multicolumn{1}{r}{(0.2)} &
  \multicolumn{1}{r}{(0.2)} &
  \multicolumn{1}{r}{(0.3)} &
  \multicolumn{1}{r}{(0.3)} &
  \multicolumn{1}{r}{(0.3)} &
  \multicolumn{1}{r}{(0.3)} &
  \multicolumn{1}{r}{(0.4)} \\
\multicolumn{1}{l}{Asian} &
  \multicolumn{1}{|r}{1.2***} &
  \multicolumn{1}{r}{1.1***} &
  \multicolumn{1}{r}{1.7***} &
  \multicolumn{1}{r}{1.9***} &
  \multicolumn{1}{r}{1.4***} &
  \multicolumn{1}{r}{3.4***} &
  \multicolumn{1}{r}{2.3***} &
  \multicolumn{1}{r}{2.7***} &
  \multicolumn{1}{r}{2.6***} &
  \multicolumn{1}{r}{1.7***} \\
\multicolumn{1}{l}{} &
  \multicolumn{1}{|r}{(0.0)} &
  \multicolumn{1}{r}{(0.0)} &
  \multicolumn{1}{r}{(0.0)} &
  \multicolumn{1}{r}{(0.0)} &
  \multicolumn{1}{r}{(0.0)} &
  \multicolumn{1}{r}{(0.1)} &
  \multicolumn{1}{r}{(0.0)} &
  \multicolumn{1}{r}{(0.0)} &
  \multicolumn{1}{r}{(0.0)} &
  \multicolumn{1}{r}{(0.1)} \\
\multicolumn{1}{l}{Black/African American} &
  \multicolumn{1}{|r}{10.2***} &
  \multicolumn{1}{r}{4.2***} &
  \multicolumn{1}{r}{3.9***} &
  \multicolumn{1}{r}{4.3***} &
  \multicolumn{1}{r}{3.6***} &
  \multicolumn{1}{r}{8.4***} &
  \multicolumn{1}{r}{4.8***} &
  \multicolumn{1}{r}{3.2***} &
  \multicolumn{1}{r}{3.2***} &
  \multicolumn{1}{r}{2.7***} \\
\multicolumn{1}{l}{} &
  \multicolumn{1}{|r}{(0.0)} &
  \multicolumn{1}{r}{(0.0)} &
  \multicolumn{1}{r}{(0.0)} &
  \multicolumn{1}{r}{(0.0)} &
  \multicolumn{1}{r}{(0.0)} &
  \multicolumn{1}{r}{(0.1)} &
  \multicolumn{1}{r}{(0.1)} &
  \multicolumn{1}{r}{(0.1)} &
  \multicolumn{1}{r}{(0.1)} &
  \multicolumn{1}{r}{(0.1)} \\
\multicolumn{1}{l}{Native Hawaiian/Pac. Islander} &
  \multicolumn{1}{|r}{7.1***} &
  \multicolumn{1}{r}{3.0***} &
  \multicolumn{1}{r}{3.5***} &
  \multicolumn{1}{r}{3.5***} &
  \multicolumn{1}{r}{2.7***} &
  \multicolumn{1}{r}{5.2***} &
  \multicolumn{1}{r}{2.5***} &
  \multicolumn{1}{r}{2.7***} &
  \multicolumn{1}{r}{2.7***} &
  \multicolumn{1}{r}{2.5***} \\
\multicolumn{1}{l}{} &
  \multicolumn{1}{|r}{(0.3)} &
  \multicolumn{1}{r}{(0.3)} &
  \multicolumn{1}{r}{(0.2)} &
  \multicolumn{1}{r}{(0.3)} &
  \multicolumn{1}{r}{(0.3)} &
  \multicolumn{1}{r}{(0.4)} &
  \multicolumn{1}{r}{(0.4)} &
  \multicolumn{1}{r}{(0.3)} &
  \multicolumn{1}{r}{(0.4)} &
  \multicolumn{1}{r}{(0.5)} \\
\multicolumn{1}{l}{Hispanic/Latino} &
  \multicolumn{1}{|r}{4.9***} &
  \multicolumn{1}{r}{0.4***} &
  \multicolumn{1}{r}{1.7***} &
  \multicolumn{1}{r}{1.9***} &
  \multicolumn{1}{r}{1.9***} &
  \multicolumn{1}{r}{4.6***} &
  \multicolumn{1}{r}{1.4***} &
  \multicolumn{1}{r}{2.0***} &
  \multicolumn{1}{r}{1.9***} &
  \multicolumn{1}{r}{1.6***} \\
\multicolumn{1}{l}{} &
  \multicolumn{1}{|r}{(0.0)} &
  \multicolumn{1}{r}{(0.0)} &
  \multicolumn{1}{r}{(0.0)} &
  \multicolumn{1}{r}{(0.0)} &
  \multicolumn{1}{r}{(0.0)} &
  \multicolumn{1}{r}{(0.0)} &
  \multicolumn{1}{r}{(0.0)} &
  \multicolumn{1}{r}{(0.0)} &
  \multicolumn{1}{r}{(0.0)} &
  \multicolumn{1}{r}{(0.1)} \\
\multicolumn{1}{l}{Multiple Responses} &
  \multicolumn{1}{|r}{4.0***} &
  \multicolumn{1}{r}{1.9***} &
  \multicolumn{1}{r}{1.9***} &
  \multicolumn{1}{r}{2.0***} &
  \multicolumn{1}{r}{1.0***} &
  \multicolumn{1}{r}{2.8***} &
  \multicolumn{1}{r}{1.6***} &
  \multicolumn{1}{r}{1.3***} &
  \multicolumn{1}{r}{1.2***} &
  \multicolumn{1}{r}{0.8***} \\
\multicolumn{1}{l}{} &
  \multicolumn{1}{|r}{(0.0)} &
  \multicolumn{1}{r}{(0.0)} &
  \multicolumn{1}{r}{(0.0)} &
  \multicolumn{1}{r}{(0.0)} &
  \multicolumn{1}{r}{(0.0)} &
  \multicolumn{1}{r}{(0.0)} &
  \multicolumn{1}{r}{(0.0)} &
  \multicolumn{1}{r}{(0.0)} &
  \multicolumn{1}{r}{(0.0)} &
  \multicolumn{1}{r}{(0.1)} \\
\multicolumn{1}{l}{Missing} &
  \multicolumn{1}{|r}{4.9***} &
  \multicolumn{1}{r}{4.0***} &
  \multicolumn{1}{r}{3.8***} &
  \multicolumn{1}{r}{3.9***} &
  \multicolumn{1}{r}{2.0***} &
  \multicolumn{1}{r}{3.2***} &
  \multicolumn{1}{r}{2.7***} &
  \multicolumn{1}{r}{2.5***} &
  \multicolumn{1}{r}{2.4***} &
  \multicolumn{1}{r}{1.2***} \\
\multicolumn{1}{l}{} &
  \multicolumn{1}{|r}{(0.0)} &
  \multicolumn{1}{r}{(0.0)} &
  \multicolumn{1}{r}{(0.0)} &
  \multicolumn{1}{r}{(0.0)} &
  \multicolumn{1}{r}{(0.0)} &
  \multicolumn{1}{r}{(0.0)} &
  \multicolumn{1}{r}{(0.0)} &
  \multicolumn{1}{r}{(0.0)} &
  \multicolumn{1}{r}{(0.0)} &
  \multicolumn{1}{r}{(0.0)} \\
\multicolumn{1}{l}{Controls} &
  \multicolumn{1}{|r}{None} &
  \multicolumn{1}{r}{Linear} &
  \multicolumn{1}{r}{Bins} &
  \multicolumn{1}{r}{Max} &
  \multicolumn{1}{r}{Lender} &
  \multicolumn{1}{r}{None} &
  \multicolumn{1}{r}{Linear} &
  \multicolumn{1}{r}{Bins} &
  \multicolumn{1}{r}{Max} &
  \multicolumn{1}{r}{Lender} \\
\multicolumn{1}{l}{Observations} &
  \multicolumn{1}{|r}{11,042,795} &
  \multicolumn{1}{r}{11,042,795} &
  \multicolumn{1}{r}{11,042,795} &
  \multicolumn{1}{r}{11,042,795} &
  \multicolumn{1}{r}{11,042,795} &
  \multicolumn{1}{r}{4,124,257} &
  \multicolumn{1}{r}{4,124,257} &
  \multicolumn{1}{r}{4,124,257} &
  \multicolumn{1}{r}{4,124,257} &
  \multicolumn{1}{r}{4,124,257} \\
\multicolumn{1}{l}{Adjusted R-squared} &
  \multicolumn{1}{|r}{0.01} &
  \multicolumn{1}{r}{0.09} &
  \multicolumn{1}{r}{0.19} &
  \multicolumn{1}{r}{0.23} &
  \multicolumn{1}{r}{0.34} &
  \multicolumn{1}{r}{0.01} &
  \multicolumn{1}{r}{0.07} &
  \multicolumn{1}{r}{0.24} &
  \multicolumn{1}{r}{0.25} &
  \multicolumn{1}{r}{0.34} \\
\cline{1-11}
\end{tabular}

\footnotesize{
Note: Each column reports results from an OLS regression of the binary variable \emph{denial} on race/ethnicity indicators and controls as indicated in the "Controls" row. Coefficients are reported in percentage points. The excluded race/ethnicity category is non-Hispanic White. Race/ethnicity categories are defined in Appendix Section \ref{sec:racecat}. Controls, when present, consist of credit score, DTI, and CLTV variables. "Linear" indicates these controls enter linearly; "Bins" indicates they enter as a series of indicators for each dimension (10 point ranges for credit score and 5 point ranges for DTI and CLTV); "Max" indicates the controls are all interactions of the indicators used in "Bins"; "Lender" indicates the controls are all interactions of the controls in "Bins," plus lender indicators. These regressions are run on two different populations defined in the text: (a) all applications excluding HRB loans and (b) standard purchase applications.
}
\end{table}

\begin{table}[!h]
	\caption{Interest Rate (basis points), Relative to Non-Hispanic White \label{tab:ratereg}}
	\centering
	\begin{tabular}{lllllllllll}
		\cline{1-11}
		\multicolumn{1}{c}{} &
		\multicolumn{5}{|c}{All Excluding HRB} &
		\multicolumn{5}{c}{Standard Purchase} \\
		\multicolumn{1}{c}{} &
		\multicolumn{1}{|r}{(1)} &
		\multicolumn{1}{r}{(2)} &
		\multicolumn{1}{r}{(3)} &
		\multicolumn{1}{r}{(4)} &
		\multicolumn{1}{r}{(5)} &
		\multicolumn{1}{r}{(1)} &
		\multicolumn{1}{r}{(2)} &
		\multicolumn{1}{r}{(3)} &
		\multicolumn{1}{r}{(4)} &
		\multicolumn{1}{r}{(5)} \\
		\cline{1-11}
		\multicolumn{1}{l}{Am. Indian/Alaska Native} &
		\multicolumn{1}{|r}{11.8***} &
		\multicolumn{1}{r}{1.6**} &
		\multicolumn{1}{r}{1.8***} &
		\multicolumn{1}{r}{1.2*} &
		\multicolumn{1}{r}{0.1} &
		\multicolumn{1}{r}{2.3***} &
		\multicolumn{1}{r}{-3.4***} &
		\multicolumn{1}{r}{-4.3***} &
		\multicolumn{1}{r}{-4.5***} &
		\multicolumn{1}{r}{0.2} \\
		\multicolumn{1}{l}{} &
		\multicolumn{1}{|r}{(0.6)} &
		\multicolumn{1}{r}{(0.6)} &
		\multicolumn{1}{r}{(0.6)} &
		\multicolumn{1}{r}{(0.6)} &
		\multicolumn{1}{r}{(0.6)} &
		\multicolumn{1}{r}{(0.8)} &
		\multicolumn{1}{r}{(0.8)} &
		\multicolumn{1}{r}{(0.8)} &
		\multicolumn{1}{r}{(0.8)} &
		\multicolumn{1}{r}{(0.8)} \\
		\multicolumn{1}{l}{Asian} &
		\multicolumn{1}{|r}{-22.8***} &
		\multicolumn{1}{r}{-16.4***} &
		\multicolumn{1}{r}{-16.1***} &
		\multicolumn{1}{r}{-15.5***} &
		\multicolumn{1}{r}{-12.0***} &
		\multicolumn{1}{r}{-12.9***} &
		\multicolumn{1}{r}{-11.9***} &
		\multicolumn{1}{r}{-10.4***} &
		\multicolumn{1}{r}{-10.3***} &
		\multicolumn{1}{r}{-8.3***} \\
		\multicolumn{1}{l}{} &
		\multicolumn{1}{|r}{(0.1)} &
		\multicolumn{1}{r}{(0.1)} &
		\multicolumn{1}{r}{(0.1)} &
		\multicolumn{1}{r}{(0.1)} &
		\multicolumn{1}{r}{(0.1)} &
		\multicolumn{1}{r}{(0.1)} &
		\multicolumn{1}{r}{(0.1)} &
		\multicolumn{1}{r}{(0.1)} &
		\multicolumn{1}{r}{(0.1)} &
		\multicolumn{1}{r}{(0.1)} \\
		\multicolumn{1}{l}{Black/African American} &
		\multicolumn{1}{|r}{18.4***} &
		\multicolumn{1}{r}{0.3***} &
		\multicolumn{1}{r}{1.0***} &
		\multicolumn{1}{r}{2.0***} &
		\multicolumn{1}{r}{3.1***} &
		\multicolumn{1}{r}{17.8***} &
		\multicolumn{1}{r}{4.3***} &
		\multicolumn{1}{r}{0.2} &
		\multicolumn{1}{r}{0.4**} &
		\multicolumn{1}{r}{1.0***} \\
		\multicolumn{1}{l}{} &
		\multicolumn{1}{|r}{(0.1)} &
		\multicolumn{1}{r}{(0.1)} &
		\multicolumn{1}{r}{(0.1)} &
		\multicolumn{1}{r}{(0.1)} &
		\multicolumn{1}{r}{(0.1)} &
		\multicolumn{1}{r}{(0.2)} &
		\multicolumn{1}{r}{(0.2)} &
		\multicolumn{1}{r}{(0.2)} &
		\multicolumn{1}{r}{(0.2)} &
		\multicolumn{1}{r}{(0.2)} \\
		\multicolumn{1}{l}{Native Hawaiian/Pac. Islander} &
		\multicolumn{1}{|r}{5.9***} &
		\multicolumn{1}{r}{-2.3***} &
		\multicolumn{1}{r}{-1.1*} &
		\multicolumn{1}{r}{-1.0} &
		\multicolumn{1}{r}{-1.0} &
		\multicolumn{1}{r}{10.3***} &
		\multicolumn{1}{r}{3.4***} &
		\multicolumn{1}{r}{2.8***} &
		\multicolumn{1}{r}{2.5***} &
		\multicolumn{1}{r}{-0.3} \\
		\multicolumn{1}{l}{} &
		\multicolumn{1}{|r}{(0.7)} &
		\multicolumn{1}{r}{(0.7)} &
		\multicolumn{1}{r}{(0.7)} &
		\multicolumn{1}{r}{(0.7)} &
		\multicolumn{1}{r}{(0.7)} &
		\multicolumn{1}{r}{(1.0)} &
		\multicolumn{1}{r}{(0.9)} &
		\multicolumn{1}{r}{(0.9)} &
		\multicolumn{1}{r}{(0.9)} &
		\multicolumn{1}{r}{(0.9)} \\
		\multicolumn{1}{l}{Hispanic/Latino} &
		\multicolumn{1}{|r}{17.7***} &
		\multicolumn{1}{r}{5.6***} &
		\multicolumn{1}{r}{6.8***} &
		\multicolumn{1}{r}{7.2***} &
		\multicolumn{1}{r}{2.6***} &
		\multicolumn{1}{r}{21.2***} &
		\multicolumn{1}{r}{11.0***} &
		\multicolumn{1}{r}{9.9***} &
		\multicolumn{1}{r}{9.6***} &
		\multicolumn{1}{r}{2.0***} \\
		\multicolumn{1}{l}{} &
		\multicolumn{1}{|r}{(0.1)} &
		\multicolumn{1}{r}{(0.1)} &
		\multicolumn{1}{r}{(0.1)} &
		\multicolumn{1}{r}{(0.1)} &
		\multicolumn{1}{r}{(0.1)} &
		\multicolumn{1}{r}{(0.1)} &
		\multicolumn{1}{r}{(0.1)} &
		\multicolumn{1}{r}{(0.1)} &
		\multicolumn{1}{r}{(0.1)} &
		\multicolumn{1}{r}{(0.1)} \\
		\multicolumn{1}{l}{Multiple Responses} &
		\multicolumn{1}{|r}{-2.0***} &
		\multicolumn{1}{r}{-7.1***} &
		\multicolumn{1}{r}{-6.4***} &
		\multicolumn{1}{r}{-6.0***} &
		\multicolumn{1}{r}{-3.5***} &
		\multicolumn{1}{r}{0.9***} &
		\multicolumn{1}{r}{-3.3***} &
		\multicolumn{1}{r}{-3.1***} &
		\multicolumn{1}{r}{-3.1***} &
		\multicolumn{1}{r}{-2.9***} \\
		\multicolumn{1}{l}{} &
		\multicolumn{1}{|r}{(0.1)} &
		\multicolumn{1}{r}{(0.1)} &
		\multicolumn{1}{r}{(0.1)} &
		\multicolumn{1}{r}{(0.1)} &
		\multicolumn{1}{r}{(0.1)} &
		\multicolumn{1}{r}{(0.1)} &
		\multicolumn{1}{r}{(0.1)} &
		\multicolumn{1}{r}{(0.1)} &
		\multicolumn{1}{r}{(0.1)} &
		\multicolumn{1}{r}{(0.1)} \\
		\multicolumn{1}{l}{Missing} &
		\multicolumn{1}{|r}{-10.4***} &
		\multicolumn{1}{r}{-10.0***} &
		\multicolumn{1}{r}{-9.5***} &
		\multicolumn{1}{r}{-9.3***} &
		\multicolumn{1}{r}{-5.0***} &
		\multicolumn{1}{r}{-4.4***} &
		\multicolumn{1}{r}{-4.6***} &
		\multicolumn{1}{r}{-4.5***} &
		\multicolumn{1}{r}{-4.5***} &
		\multicolumn{1}{r}{-3.7***} \\
		\multicolumn{1}{l}{} &
		\multicolumn{1}{|r}{(0.1)} &
		\multicolumn{1}{r}{(0.1)} &
		\multicolumn{1}{r}{(0.1)} &
		\multicolumn{1}{r}{(0.1)} &
		\multicolumn{1}{r}{(0.1)} &
		\multicolumn{1}{r}{(0.1)} &
		\multicolumn{1}{r}{(0.1)} &
		\multicolumn{1}{r}{(0.1)} &
		\multicolumn{1}{r}{(0.1)} &
		\multicolumn{1}{r}{(0.1)} \\
		\multicolumn{1}{l}{Controls} &
		\multicolumn{1}{|r}{None} &
		\multicolumn{1}{r}{Linear} &
		\multicolumn{1}{r}{Bins} &
		\multicolumn{1}{r}{Max} &
		\multicolumn{1}{r}{Lender} &
		\multicolumn{1}{r}{None} &
		\multicolumn{1}{r}{Linear} &
		\multicolumn{1}{r}{Bins} &
		\multicolumn{1}{r}{Max} &
		\multicolumn{1}{r}{Lender} \\
		\multicolumn{1}{l}{Observations} &
		\multicolumn{1}{|r}{9,683,087} &
		\multicolumn{1}{r}{9,683,087} &
		\multicolumn{1}{r}{9,683,087} &
		\multicolumn{1}{r}{9,683,087} &
		\multicolumn{1}{r}{9,683,087} &
		\multicolumn{1}{r}{3,832,227} &
		\multicolumn{1}{r}{3,832,227} &
		\multicolumn{1}{r}{3,832,227} &
		\multicolumn{1}{r}{3,832,227} &
		\multicolumn{1}{r}{3,832,227} \\
		\multicolumn{1}{l}{Adjusted R-squared} &
		\multicolumn{1}{|r}{0.02} &
		\multicolumn{1}{r}{0.09} &
		\multicolumn{1}{r}{0.11} &
		\multicolumn{1}{r}{0.13} &
		\multicolumn{1}{r}{0.36} &
		\multicolumn{1}{r}{0.02} &
		\multicolumn{1}{r}{0.13} &
		\multicolumn{1}{r}{0.16} &
		\multicolumn{1}{r}{0.17} &
		\multicolumn{1}{r}{0.32} \\
		\cline{1-11}
	\end{tabular}
	
	\footnotesize{
		Note: Each column reports results from an OLS regression of interest rate on race/ethnicity indicators and controls as indicated in the "Controls" row. Coefficients are reported in percentage points. The excluded race/ethnicity category is non-Hispanic White. Race/ethnicity categories are defined in Appendix Section \ref{sec:racecat}. Controls, when present, consist of credit score, DTI, and CLTV variables. "Linear" indicates these controls enter linearly; "Bins" indicates they enter as a series of indicators for each dimension (10 point ranges for credit score and 5 point ranges for DTI and CLTV); "Max" indicates the controls are all interactions of the indicators used in "Bins"; "Lender" indicates the controls are all interactions of the controls in "Bins," plus lender indicators. These regressions are run on two different populations defined in the text: (a) all applications excluding HRB loans and (b) standard purchase applications.
	}
\end{table}
	\end{landscape}
}	
	Table \ref{tab:denialreg} shows regression results for denial probability. The columns without controls, labeled (1), show unconditional differences in denial rates. For American Indian/Alaska Native and Black applicants, the disparities (8.7 and 10.2 percent, respectively) are especially high when looking at all applications excluding HRB loans. Non-Hispanic White applicants have just over a 10 percent chance of denial in this population, meaning that Black applicants are about twice as likely to be denied.\footnote{As our focus is studying aggregate disparities, we include applications with low credit score or high LTV or DTI, in contrast to \textcite{kylim2022} who exclude these applications. Accordingly, the overall denial rate in our sample is higher.} Hispanic applicants, with a 4.9 percentage point higher chance of denial, are about 50 percent more likely to be denied than non-Hispanic White applicants. 
	
	Asian applicants have the smallest (but still statistically significant) unconditional disparities relative to non-Hispanic White applicants. Multiple-response applicants and applicants with no race or ethnicity response face moderate disparities relative to the other racial and ethnic groups, suggesting these at least some of these applicants may face the same downward forces on access to credit as the minorities that we are able to identify. While the percentage point disparities for standard purchase applications are smaller, the non-Hispanic White denial rate is also lower---5.3 percent. Therefore, for these loans, Black applicants' 8.4 percentage point disparity translates to a denial rate over 1.5 times higher than that of non-Hispanic White applicants.
	
	As we add creditworthiness controls in columns (2)-(4), we observe disparities drop. These results are consistent with these creditworthiness factors being negatively correlated with minority status. In other words, part of the reason that minorities have higher denial rates is because, on average, creditworthiness indicators of minority applicants are more likely to fall below mortgage eligibility cutoffs. The linear controls significantly improve model fit, as measured by the adjusted R-squared. The binned controls improved fit further, but interacting the bins only slightly improves model fit and has negligible effect on the measured disparities.
	
	Including lender fixed effects narrows the estimated disparities further. This result suggests that  lenders who tend to serve applicants from minority racial and ethnic groups may have higher underwriting standards. It may also be the case that borrowers sort into lenders partially based on creditworthiness factors that we do not observe and that are negatively correlated with minority status. In either case, these patterns may present an aggregate barrier to credit for those groups even if each lender is treating applicants fairly. 
	
	Though disparities fall with creditworthiness controls, the conditional disparities remain economically significant. Black applicants for standard purchase loans are more than half again as likely to be denied as non-Hispanic White applicants with the same credit score, DTI, and CLTV combination, at the same lender. Hispanic applicants for standard purchase loans are about 30 percent more likely to face denial than non-Hispanic White applicants with the same credit score, DTI, and CLTV combination at the same lender, and this is the smallest disparity among the groups we studied.
	
	From the amounts disparities drop with the addition of controls, it is evident that the three creditworthiness controls may themselves be significant drivers of unconditional disparities. Since our regressions do not include other creditworthiness factors that are likely correlated with the three included controls as well as with denial, these results do not necessarily reflect the extent to which these three factors account for disparities in underwriting decisions. Disparities in actual repayment rates, which may result from a variety of structural economic differences such as the racial wage gap, may limit the extent to which disparities in mortgage outcomes can be reduced by replacing these variables with alternative measures of creditworthiness. 
	
	Table \ref{tab:ratereg} shows regression results for interest rate. Unconditional disparities follow similar patterns to denial disparities. Black borrowers face the largest price disparity at over 18 basis points for all loans excluding HRB. While this is about a 4.2 percent higher interest rate than the non-Hispanic White average of 4.4 percent, because interest compounds, this translates to about 4.9   percent more interest paid over the course of a 30-year mortgage (holding loan size fixed). 
	
	After controlling for creditworthiness indicators, Hispanic borrowers' disparities fall by more than half, and those for Black borrowers fall to two basis points. Incorporating lender fixed effects reduces the disparities for Hispanic borrowers to less than three basis points as well. Across the board, Asian borrowers have significantly lower interest rates than non-Hispanic White borrowers. This may be the result of Asian borrowers paying discount points more often, which we do not account for in these regressions. The remaining disparities do not seem to owe to some groups tending to borrow when rates are higher--we observed no appreciable racial disparities in the national average 30-year mortgage rate (at the time of loan decision) for our sample.
	
	The Appendix contains two robustness checks for our pricing regressions. Table \ref{tab:APRspread} shows the same regressions as in Table \ref{tab:ratereg} for Standard Home Purchases, but replaces interest rate as the outcome of interest with lender-reported APR spread. APR spread is defined as the difference in Annual Percentage Rate (a measure of loan costs that includes both interest rate and certain up-front costs) between the loan in question and a comparable transaction in terms of amortization and term from the FFIEC website.\footnote{See 12 CFR § 1003.4(a)(12)(i) for details.} The results show racial and ethnic disparities in APR rate spread comparable to those for interest rates, after controlling for binned credit score, DTI, and CLTV as well as lender fixed effects. This suggests that the observed disparities in interest rates are not offset by differences in up-front costs (which APR includes). 
	
	To investigate whether the observed interest rate disparities owe to differences in discount points in particular (which borrowers can elect to pay up-front to lower their interest rate) Table \ref{tab:StdHPnodisc} repeats the interest rate regressions for Standard Purchases from Table \ref{tab:ratereg}, but restricting the sample to Standard Purchase loans for which the borrowers had zero discount points. The results are attenuated somewhat but still qualitatively similar, demonstrating that interest rate disparities exist even for borrowers who chose zero discount points.\footnote{As \cite{zhang2021lenders} show, however, restricting the sample to borrowers who chose zero discount points cannot rule out the possibility that they were offered different menus (tradeoffs between interest rate and discount points).} 
	
	\section{Sensitivity analysis}\label{sec:sensitivity}
	The regression results presented in Section \ref{sec:regressions} do not include all application characteristics that lenders may be using for underwriting and pricing. If one such factor is correlated with race or ethnicity, this may account (in whole or in part) for the racial and ethnic disparities suggested by the results presented in Section \ref{sec:regressions}. In other words, if lenders use an additional variable $z$ (such as cash reserves) that is correlated with race or ethnicity as well as our outcomes of interest (denial/pricing), it may be the case that our estimated coefficients on race/ethnicity would be much smaller, would be zero, or would change signs if we were able to include $z$ in the regressions (classic omitted variable bias). 
	
	One approach is to use proxies for omitted variables. However, in this context, available proxies may not capture the racial/ethnic disparities in application characteristics used by lenders. For example, some lenders use cash reserves in underwriting. While the HMDA data include applicant income, which can be used as a proxy for wealth (as in \textcite{kylim2022} or \textcite{popick2022did}), stark racial inequalities in wealth exist even holding fixed income---see, for example, \textcite{brookings2020wealth}. Accordingly, using income as a proxy for wealth may fail to capture racial inequality in wealth and thus overstate the degree to which racial disparities in pricing or underwriting remain unexplained. In this section, we measure the sensitivity of our results to potential omitted variable bias using techniques developed by \textcite{cinelli2020making}. 
	
	Consider a variable $z$ that is used by lenders in underwriting or pricing but not included in our regressions, and suppose it is also associated with race or ethnicity. We ask how strong a predictor such a variable would have to be to fully account for (nullify) the coefficient on a given prohibited basis group indicator in Section \ref{sec:regressions}. The rightmost column in Table \ref{tab:sensitivity}, called the ``robustness value'' by \textcite{cinelli2020making}, shows the answer by racial/ethnic group for the estimated coefficients from Section \ref{sec:regressions}'s specification (3).\footnote{Specification 3 (including binned variables but not their interactions) was chosen for sensitivity analysis as it is the richest specification for which the predictive variables are separable, which permits benchmarking a potential confounder variable to the predictive power of credit score (see $\dagger$ in Table \ref{tab:sensitivity}).}
	
	The interpretation of the robustness value (RV) is as follows. Looking at the third-to-last row, for example, an omitted determinant of denial (orthogonal to the covariates) that explains at least 2.29 percent of the residual variance\footnote{The ``residual variance’’ of a variable (either the PBG indicator or the outcome variable) is the variation unexplained by existing control variables (e.g., the amount of residual variance of outcome $y$ explained by $z$ after controlling for $X$ and $pbg$ is $R^2_{y\sim z \vert X,pbg} = \frac{R^2_{y\sim z+X+pbg}-R^2_{y\sim X+pbg}}{1-R^2_{y\sim X+pbg}}$) .} of both the Hispanic/Latino indicator   and the outcome (denial) is strong enough to bring the point estimate of the coefficient on this PBG indicator to 0 (a bias of 100 percent of the original estimate). In other words, $z$ would be sufficient to nullify the coefficient on that PBG estimated in Section \ref{sec:regressions} if the association of $z$ with both the outcome (denial, here) and the prohibited basis group were at least equal to the robustness value (that is, $R^2_{y \sim z \vert X, pbg}, R^2_{pbg\sim z \vert X} >RV$, where $X$ are the observable factors we use as controls).\footnote{While similar in spirit to the sensitivity analysis proposed by \textcite{aet2005}, this is an assumption about the additional predictive power provided by $z$, rather than their assumption that ``the part of [the outcome] that is related to the observables and the part related to the unobservables have the same relationship with [the treatment].''}

	\begin{table}[!h]
		\caption{Sensitivity of results to omitted variables\label{tab:sensitivity}}
		Outcome: Interest rate \\
		\\
		\begin{tabular}{l|cc}
			\hline
			Racial/ethnic group& Partial $R^2$ of PBG with interest rate & Robustness Value \\
			\hline 
			American Indian/Alaska Native& 0.00\% & 0.31\%$^{\dagger}$ \\
			
			Asian& 0.23\%  & 4.65\%$^{\dagger\dagger\dagger}$ \\
			
			Black or African American& 0.00\% & 0.06\% \\
			
			Native Hawaiian/Pacific Islander& 0.00\% & 0.17\% \\
			
			Hispanic/Latino& 0.23\% & 4.71\%$^{\dagger}$ \\
			
			Multiple responses& 0.02\% & 1.49\% \\
			
			Missing& 0.06\% & 2.48\%$^{\dagger\dagger\dagger}$ \\
			
			\hline
		\end{tabular}
		\\
		\\
		\\
		Outcome: Denial \\
		\\
		\begin{tabular}{l|cc}
			\hline
			Racial/ethnic group& Partial $R^2$ of PBG with denial & Robustness Value \\
			\hline 
			American Indian/Alaska Native& 0.00\% & 0.33\%$^{\dagger\dagger\dagger}$ \\
			
			Asian& 0.09\%  & 2.93\%$^{\dagger\dagger\dagger}$ \\
			
			Black or African American& 0.09\% & 2.90\%$^{\dagger}$ \\
			
			Native Hawaiian/Pacific Islander& 0.00\% & 0.40\%$^{\dagger\dagger}$ \\
			
			Hispanic/Latino& 0.05\% & 2.29\%$^{\dagger}$ \\
			
			Multiple responses& 0.02\% & 1.48\%$^{\dagger}$ \\
			
			Missing& 0.11\% & 3.21\%$^{\dagger\dagger\dagger}$ \\ 
			\hline
			
		\end{tabular}
		\\
		\begin{minipage}{0.95\textwidth}
			\begin{singlespace}
				
				{\footnotesize Note: Robustness Value is the minimum value of $R^2_{y \sim z \vert X, pbg}$ and $R^2_{pbg\sim z \vert X}$ sufficient for an omitted variable $z$ to fully explain (nullify) the coefficient on $pbg$ in a regression of the outcome $y$ on $X$ and $pbg$. Partial $R^2$ of PBG with the outcome is $R^2_{y\sim pbg\vert X}$. The variables $X$ are as in specification (3) in Section \ref{sec:regressions}. $^{\dagger}$, $^{\dagger\dagger}$, and $^{\dagger\dagger\dagger}$ denote PBG coefficients that are robust to (could not be nullified by) an omitted confounder $1\times$, $2\times$, and $3\times$ as strong as credit score, respectively. Computed using \emph{sensemakr} (\cite{cinelli2020sensemakr}.}
			\end{singlespace}
		\end{minipage}
	\end{table}
	How plausible are these values? Is it reasonable that an omitted confounding variable might have sufficient additional predictive power for both the outcome and prohibited basis group membership (beyond the predictive power of our three key variables)? For reference, take credit score (which is included in our regressions) as a benchmark. Consider an omitted confounding variable $z$ that has the same additional predictive power as credit score on the outcome as well as the same association with a given prohibited basis group.\footnote{For example, consider a potential confounder $z$ that has the same additional predictive power for denial and for Hispanic/Latino as does credit score. Such a variable $z$ would have $R^2_{y\sim z\vert pbg,X}=4.52\%$ and $R^2_{pbg\sim z\vert X}=0.74\%$ (the strength of credit score is measured as its additional predictive power beyond that of the other covariates---see \textcite{cinelli2020making} Section 4.4 for details). Note that the association with the outcome is above the Robustness Value (2.29\%) while the association with the PBG is too low, so the Robustness Value is inconclusive here. However, we can also consider other (unequal) combinations of $R^2_{y\sim z\vert pbg,X}$ and $R^2_{pbg\sim z\vert X}$. As it turns out, such a $z$ would be insufficiently strong to nullify the estimated coefficient on Hispanic/Latino ethnicity, but one twice as strong (in terms of association with outcome and PBG) would be, as denoted by the $\dagger$ in Table \ref{tab:sensitivity}.}
	
	For the interest rate regressions, such confounders would be strong enough to nullify the estimated coefficients for the Black/African American, Native Hawaiian/Pacific Islander, and Multiple Responses groups. And omitted confounders twice as strong\footnote{That is, with twice the predictive power for both outcome and PBG indicator.} as credit score would be sufficient to nullify the coefficients for American Indian/Alaska Native and Hispanic/Latino, but confounders as strong as credit score would not be sufficient (distinguished by $^{\dagger}$ in Table \ref{tab:sensitivity}). For the denial regressions, confounders twice as strong as credit score would be sufficient to nullify the coefficients on Black/African American, Hispanic/Latino, and Multiple Responses (but not confounders as strong as credit score), and confounders three times as strong as credit score would suffice to nullify the coefficient on Native Hawaiian/Pacific Islander (but not confounders twice as strong as credit score).
	
	\textcite{cinelli2020making} provide another way of thinking about the sensitivity of the estimated coefficients as well. Suppose an omitted variable $z$ fully explains the residual variation in outcomes. How correlated would such a variable have to be with race/ethnicity to nullify (reduce to zero) the coefficients estimated in Table \ref{sec:regressions}? \textcite{cinelli2020making} show that the answer is the partial $R^2$ of the race/ethnicity indicator with the outcome, given in the middle column of Table \ref{tab:sensitivity} for each racial/ethnic group. Taking again the third-to-last row as an example, a confounder variable (orthogonal to the covariates) that explains all of the residual variance of the outcome would need to explain at least 0.05 percent of the residual variance of the Hispanic/Latino indicator to fully account for the observed estimated coefficient on this PBG indicator. 
	
	Taken together, these sensitivity analyses caution that the results of Section \ref{sec:regressions} may owe to variables that determine underwriting or pricing outcomes but that are not included in our regressions, especially if those variables are strongly correlated with race or ethnicity.
	
	\section{Conclusion}\label{sec:discussion}
	This paper uses 2018-2019 HMDA mortgage application data to study disparities in outcomes. We first show how application characteristics (credit score, DTI, and CLTV) differ by race and ethnicity and how application denial and interest rates vary with each characteristic individually. These descriptive exercises suggest that these three factors play an important role in racial and ethnic disparities in the mortgage market. Using regression analysis, we show that this finding is robust to simultaneously controlling for the three application characteristics. However, these factors together are unable to fully explain racial and ethnic disparities in mortgage approval or interest rates. The raw disparities in denial rates are, for the racial and ethnic groups studied, attenuated by roughly half once controls are added (Table \ref{tab:denialreg}). For interest rates, the results are much more mixed: Controls are able to account for raw disparities for some groups but not for others. Our sensitivity analysis indicates that omitted factors that are predictive of approval or of pricing and of race/ethnicity to the same extent as credit score are not sufficient to explain the remaining disparities for several racial and ethnic groups. 
	
	While our methodology is not able to address the question of whether there is discrimination in the mortgage market, our results nevertheless indicate that barriers to obtaining and affording a
	mortgage more frequently impact minority borrowers. Credit score, income, and down payment have become nearly universal indicators of credit risk in mortgage underwriting, driven in large part by the policies of Government-Sponsored Enterprises like Fannie Mae and Freddie Mac who purchase a substantial portion of mortgages from originators. Our data suggest use of these factors (and correlated datapoints not included) presents a particularly large obstacle for minority mortgage applicants (though no precedent suggests they are legally impermissible). The wider lending industry is currently experimenting with new measures of creditworthiness, though it remains to be seen to what extent they will be able to expand credit access.

	The disparities that remain after controlling for these three factors are similarly economically significant. The barriers that these remaining disparities represent may be additional creditworthiness, market, or collateral factors orthogonal to the three we control for. They may also be differences in financial literacy across racial and ethnic groups if more savvy applicants are less likely to fail to provide appropriate documentation or misunderstand application instructions. Illegal discrimination, to the extent it occurs in the mortgage market, could also contribute to the remaining disparities. Each of these explanations requires a different policy response, and thus continued study of these disparities is valuable to guide our society’s approach to addressing the systemic gaps in mortgage access and affordability that minority borrowers continue to face.
	
	\clearpage
	\printbibliography[]
	
	\clearpage
	
	\section{Appendix}
	\subsection{Race categories in HMDA}\label{sec:HMDArace}
	\begin{singlespace}
		The race categories used in HMDA are as follows:
		\begin{itemize}
			\setlength\itemsep{-0.5em}
			\item American Indian or Alaska Native
			\item Asian
			\begin{itemize} \setlength\itemsep{-0.2em} \vspace{-1.0em}
				\item Asian Indian
				\item Chinese
				\item Filipino
				\item Japanese
				\item Korean
				\item Vietnamese
				\item Other Asian
			\end{itemize}
			\item Black or African American
			\item Native Hawaiian or Other Pacific Islander
			\begin{itemize}\setlength\itemsep{-0.2em} \vspace{-1.0em}
				\item Native Hawaiian
				\item Guamanian or Chamorro
				\item Samoan
				\item Other Pacific Islander
			\end{itemize}
			\item White
		\end{itemize}
		
		The ethnicity categories in HMDA are:
		
		\begin{itemize}
			\setlength\itemsep{-0.5em}
			\item Hispanic or Latino
			\begin{itemize}
				\setlength\itemsep{-0.2em} \vspace{-1.0em}
				\item Mexican
				\item Puerto Rican
				\item Cuban
				\item Other Hispanic or Latino
			\end{itemize}
			\item Not Hispanic or Latino
		\end{itemize}
	\end{singlespace}
	
	\subsection{Our race and ethnicity categories}\label{sec:racecat}
	We first combine applicants' multiple responses and responses from multiple applicants into one of the mutually exclusive categories below. Where ``only'' is indicated, we require all responses to be the noted race or ethnicity. No distinction is made between applicant and co-applicant responses except where noted. ``Not Hispanic or Latino'' responses are ignored except where noted. Missing values are ignored unless all are missing, which is one of the categories below.
	
	\begin{itemize}
		\setlength\itemsep{-0.5em}
		\item American Indian only
		\item Asian Indian only (ignoring Asian if same applicant is Asian Indian)
		\item Chinese only (ignoring Asian if same applicant is Chinese)
		\item Filipino only (ignoring Asian if same applicant is Asian Indian)
		\item Japanese only (ignoring Asian if same applicant is Filipino)
		\item Korean only (ignoring Asian if same applicant is Korean)
		\item Vietnamese only (ignoring Asian if same applicant is Vietnamese)
		\item All Asian/Asian subcategories only not in previous 6 categories
		\item Black only
		\item Native Hawaiian only (ignoring Pacific Islander if same applicant is Native Hawaiian)
		\item Guamanian or Chamorro only (ignoring Pacific Islander if same applicant is Guamanian or Chamorro)
		\item Samoan only (ignoring Pacific Islander if same applicant is Samoan)
		\item All Pacific Islander/Pacific Islander subcategories only not in previous 3 categories
		\item Non-Hispanic White only
		\item Mexican only (ignoring Hispanic if same applicant is Mexican; no non-Hispanic)
		\item Puerto Rican only (ignoring Hispanic if same applicant is Puerto Rican; no non-Hispanic)
		\item Cuban only (ignoring Hispanic if same applicant is Cuban; no non-Hispanic)
		\item All Hispanic/Hispanic subcategories three previous categories (no non-Hispanic)
		\item All other combinations of reports
		\item All missing
	\end{itemize}
	
	We use these categories in Figure \ref{fig:denialraceCLTV}. For all other analyses, we aggregate the categories above into eight major categories as follows
	
	\begin{itemize}
		\setlength\itemsep{-0.5em}
		\item American Indian
		\begin{itemize}\setlength\itemsep{-0.2em} \vspace{-1.0em}
			\item American Indian only
		\end{itemize}
		\item Asian
		\begin{itemize}\setlength\itemsep{-0.2em} \vspace{-1.0em}
			\item Asian Indian only (ignoring Asian if same applicant is Asian Indian)
			\item Chinese only (ignoring Asian if same applicant is Chinese)
			\item Filipino only (ignoring Asian if same applicant is Asian Indian)
			\item Japanese only (ignoring Asian if same applicant is Filipino)
			\item Korean only (ignoring Asian if same applicant is Korean)
			\item Vietnamese only (ignoring Asian if same applicant is Vietnamese)
			\item All Asian/Asian subcategories only not in previous 6 categories
		\end{itemize}
		\item Black
		\begin{itemize}\setlength\itemsep{-0.2em} \vspace{-1.0em}	
			\item Black only
		\end{itemize}
		\item Pacific Islander
		\begin{itemize}\setlength\itemsep{-0.2em} \vspace{-1.0em}
			\item Native Hawaiian only (ignoring Pacific Islander if same applicant is Native Hawaiian)
			\item Guamanian or Chamorro only (ignoring Pacific Islander if same applicant is Guamanian or Chamorro)
			\item Samoan only (ignoring Pacific Islander if same applicant is Samoan)
			\item All Pacific Islander/Pacific Islander subcategories only not in previous 3 categories
		\end{itemize}
		\item Non-Hispanic White
		\begin{itemize}\setlength\itemsep{-0.2em} \vspace{-1.0em}
			\item Non-Hispanic White only
		\end{itemize}
		\item Hispanic
		\begin{itemize}\setlength\itemsep{-0.2em} \vspace{-1.0em}
			\item Mexican only (ignoring Hispanic if same applicant is Mexican; no non-Hispanic)
			\item Puerto Rican only (ignoring Hispanic if same applicant is Puerto Rican; no non-Hispanic)
			\item Cuban only (ignoring Hispanic if same applicant is Cuban; no non-Hispanic)
			\item All Hispanic/Hispanic subcategories three previous categories (no non-Hispanic)
		\end{itemize}
		\item Multiple Responses
		\begin{itemize}\setlength\itemsep{-0.2em} \vspace{-1.0em}
			\item All other combinations of reports
		\end{itemize}
		\item All Missing
	\end{itemize}

\subsection{Ancillary results}
\afterpage{
	\begin{landscape}
		\begin{table}[!h]
\caption{APR rate spread (basis points), Relative to Non-Hispanic White \label{tab:APRspread}}
\centering
\begin{tabular}{llllll}
\cline{1-6}
\multicolumn{1}{c}{} &
  \multicolumn{5}{|c}{Std Purchase} \\
\multicolumn{1}{c}{} &
  \multicolumn{1}{|r}{(1)} &
  \multicolumn{1}{r}{(2)} &
  \multicolumn{1}{r}{(3)} &
  \multicolumn{1}{r}{(4)} &
  \multicolumn{1}{r}{(5)} \\
\cline{1-6}
\multicolumn{1}{l}{Am. Indian/Alaska Native} &
  \multicolumn{1}{|r}{14.8***} &
  \multicolumn{1}{r}{14.8***} &
  \multicolumn{1}{r}{14.8***} &
  \multicolumn{1}{r}{2.3***} &
  \multicolumn{1}{r}{2.5***} \\
\multicolumn{1}{l}{} &
  \multicolumn{1}{|r}{(0.9)} &
  \multicolumn{1}{r}{(0.9)} &
  \multicolumn{1}{r}{(0.9)} &
  \multicolumn{1}{r}{(0.7)} &
  \multicolumn{1}{r}{(0.7)} \\
\multicolumn{1}{l}{Asian} &
  \multicolumn{1}{|r}{-18.6***} &
  \multicolumn{1}{r}{-18.6***} &
  \multicolumn{1}{r}{-18.6***} &
  \multicolumn{1}{r}{-11.2***} &
  \multicolumn{1}{r}{-7.8***} \\
\multicolumn{1}{l}{} &
  \multicolumn{1}{|r}{(0.1)} &
  \multicolumn{1}{r}{(0.1)} &
  \multicolumn{1}{r}{(0.1)} &
  \multicolumn{1}{r}{(0.1)} &
  \multicolumn{1}{r}{(0.1)} \\
\multicolumn{1}{l}{Black/African American} &
  \multicolumn{1}{|r}{29.7***} &
  \multicolumn{1}{r}{29.7***} &
  \multicolumn{1}{r}{29.7***} &
  \multicolumn{1}{r}{0.7***} &
  \multicolumn{1}{r}{3.0***} \\
\multicolumn{1}{l}{} &
  \multicolumn{1}{|r}{(0.2)} &
  \multicolumn{1}{r}{(0.2)} &
  \multicolumn{1}{r}{(0.2)} &
  \multicolumn{1}{r}{(0.2)} &
  \multicolumn{1}{r}{(0.1)} \\
\multicolumn{1}{l}{Native Hawaiian/Pac. Islander} &
  \multicolumn{1}{|r}{14.1***} &
  \multicolumn{1}{r}{14.1***} &
  \multicolumn{1}{r}{14.1***} &
  \multicolumn{1}{r}{2.8***} &
  \multicolumn{1}{r}{0.1} \\
\multicolumn{1}{l}{} &
  \multicolumn{1}{|r}{(1.0)} &
  \multicolumn{1}{r}{(1.0)} &
  \multicolumn{1}{r}{(1.0)} &
  \multicolumn{1}{r}{(0.8)} &
  \multicolumn{1}{r}{(0.7)} \\
\multicolumn{1}{l}{Hispanic/Latino} &
  \multicolumn{1}{|r}{32.6***} &
  \multicolumn{1}{r}{32.6***} &
  \multicolumn{1}{r}{32.6***} &
  \multicolumn{1}{r}{13.5***} &
  \multicolumn{1}{r}{5.2***} \\
\multicolumn{1}{l}{} &
  \multicolumn{1}{|r}{(0.1)} &
  \multicolumn{1}{r}{(0.1)} &
  \multicolumn{1}{r}{(0.1)} &
  \multicolumn{1}{r}{(0.1)} &
  \multicolumn{1}{r}{(0.1)} \\
\multicolumn{1}{l}{Multiple Responses} &
  \multicolumn{1}{|r}{3.4***} &
  \multicolumn{1}{r}{3.4***} &
  \multicolumn{1}{r}{3.4***} &
  \multicolumn{1}{r}{-2.9***} &
  \multicolumn{1}{r}{-2.1***} \\
\multicolumn{1}{l}{} &
  \multicolumn{1}{|r}{(0.1)} &
  \multicolumn{1}{r}{(0.1)} &
  \multicolumn{1}{r}{(0.1)} &
  \multicolumn{1}{r}{(0.1)} &
  \multicolumn{1}{r}{(0.1)} \\
\multicolumn{1}{l}{Missing} &
  \multicolumn{1}{|r}{-3.7***} &
  \multicolumn{1}{r}{-3.7***} &
  \multicolumn{1}{r}{-3.7***} &
  \multicolumn{1}{r}{-3.6***} &
  \multicolumn{1}{r}{-2.3***} \\
\multicolumn{1}{l}{} &
  \multicolumn{1}{|r}{(0.1)} &
  \multicolumn{1}{r}{(0.1)} &
  \multicolumn{1}{r}{(0.1)} &
  \multicolumn{1}{r}{(0.1)} &
  \multicolumn{1}{r}{(0.1)} \\
\multicolumn{1}{l}{Controls} &
  \multicolumn{1}{|r}{None} &
  \multicolumn{1}{r}{Linear} &
  \multicolumn{1}{r}{Bins} &
  \multicolumn{1}{r}{Max} &
  \multicolumn{1}{r}{Lender} \\
\multicolumn{1}{l}{Observations} &
  \multicolumn{1}{|r}{3,774,344} &
  \multicolumn{1}{r}{3,774,344} &
  \multicolumn{1}{r}{3,774,344} &
  \multicolumn{1}{r}{3,774,344} &
  \multicolumn{1}{r}{3,774,344} \\
\multicolumn{1}{l}{Adjusted R-squared} &
  \multicolumn{1}{|r}{0.04} &
  \multicolumn{1}{r}{0.04} &
  \multicolumn{1}{r}{0.04} &
  \multicolumn{1}{r}{0.42} &
  \multicolumn{1}{r}{0.62} \\
\cline{1-6}
\end{tabular}

\footnotesize{
Note: Each column reports results from an OLS regression of interest rate on race/ethnicity indicators and controls as indicated in the "Controls" row. Coefficients are reported in percentage points. The excluded race/ethnicity category is non-Hispanic White. Race/ethnicity categories are defined in Appendix Section \ref{sec:racecat}. Controls, when present, consist of credit score, DTI, and CLTV variables. "Linear" indicates these controls enter linearly; "Bins" indicates they enter as a series of indicators for each dimension (10 point ranges for credit score and 5 point ranges for DTI and CLTV); "Max" indicates the controls are all interactions of the indicators used in "Bins"; "Lender" indicates the controls are all interactions of the controls in "Bins," plus lender indicators. These regressions are run on standard purchase applications with zero discount points.
}
\end{table}

\begin{table}[!h]
\caption{Interest Rate (basis points), Relative to Non-Hispanic White \label{tab:StdHPnodisc}}
\centering
\begin{tabular}{llllll}
\cline{1-6}
\multicolumn{1}{c}{} &
  \multicolumn{5}{|c}{Std Purchase (no discount pts)} \\
\multicolumn{1}{c}{} &
  \multicolumn{1}{|r}{(1)} &
  \multicolumn{1}{r}{(2)} &
  \multicolumn{1}{r}{(3)} &
  \multicolumn{1}{r}{(4)} &
  \multicolumn{1}{r}{(5)} \\
\cline{1-6}
\multicolumn{1}{l}{Am. Indian/Alaska Native} &
  \multicolumn{1}{|r}{4.6***} &
  \multicolumn{1}{r}{4.6***} &
  \multicolumn{1}{r}{4.6***} &
  \multicolumn{1}{r}{-2.9***} &
  \multicolumn{1}{r}{1.3} \\
\multicolumn{1}{l}{} &
  \multicolumn{1}{|r}{(1.0)} &
  \multicolumn{1}{r}{(1.0)} &
  \multicolumn{1}{r}{(1.0)} &
  \multicolumn{1}{r}{(0.9)} &
  \multicolumn{1}{r}{(1.1)} \\
\multicolumn{1}{l}{Asian} &
  \multicolumn{1}{|r}{-13.4***} &
  \multicolumn{1}{r}{-13.4***} &
  \multicolumn{1}{r}{-13.4***} &
  \multicolumn{1}{r}{-10.2***} &
  \multicolumn{1}{r}{-8.6***} \\
\multicolumn{1}{l}{} &
  \multicolumn{1}{|r}{(0.1)} &
  \multicolumn{1}{r}{(0.1)} &
  \multicolumn{1}{r}{(0.1)} &
  \multicolumn{1}{r}{(0.1)} &
  \multicolumn{1}{r}{(0.2)} \\
\multicolumn{1}{l}{Black/African American} &
  \multicolumn{1}{|r}{21.6***} &
  \multicolumn{1}{r}{21.6***} &
  \multicolumn{1}{r}{21.6***} &
  \multicolumn{1}{r}{2.6***} &
  \multicolumn{1}{r}{2.5***} \\
\multicolumn{1}{l}{} &
  \multicolumn{1}{|r}{(0.2)} &
  \multicolumn{1}{r}{(0.2)} &
  \multicolumn{1}{r}{(0.2)} &
  \multicolumn{1}{r}{(0.2)} &
  \multicolumn{1}{r}{(0.2)} \\
\multicolumn{1}{l}{Native Hawaiian/Pac. Islander} &
  \multicolumn{1}{|r}{14.7***} &
  \multicolumn{1}{r}{14.7***} &
  \multicolumn{1}{r}{14.7***} &
  \multicolumn{1}{r}{5.6***} &
  \multicolumn{1}{r}{1.1} \\
\multicolumn{1}{l}{} &
  \multicolumn{1}{|r}{(1.2)} &
  \multicolumn{1}{r}{(1.2)} &
  \multicolumn{1}{r}{(1.2)} &
  \multicolumn{1}{r}{(1.1)} &
  \multicolumn{1}{r}{(1.2)} \\
\multicolumn{1}{l}{Hispanic/Latino} &
  \multicolumn{1}{|r}{25.9***} &
  \multicolumn{1}{r}{25.9***} &
  \multicolumn{1}{r}{25.9***} &
  \multicolumn{1}{r}{12.8***} &
  \multicolumn{1}{r}{3.2***} \\
\multicolumn{1}{l}{} &
  \multicolumn{1}{|r}{(0.2)} &
  \multicolumn{1}{r}{(0.2)} &
  \multicolumn{1}{r}{(0.2)} &
  \multicolumn{1}{r}{(0.2)} &
  \multicolumn{1}{r}{(0.2)} \\
\multicolumn{1}{l}{Multiple Responses} &
  \multicolumn{1}{|r}{2.1***} &
  \multicolumn{1}{r}{2.1***} &
  \multicolumn{1}{r}{2.1***} &
  \multicolumn{1}{r}{-2.1***} &
  \multicolumn{1}{r}{-2.4***} \\
\multicolumn{1}{l}{} &
  \multicolumn{1}{|r}{(0.1)} &
  \multicolumn{1}{r}{(0.1)} &
  \multicolumn{1}{r}{(0.1)} &
  \multicolumn{1}{r}{(0.1)} &
  \multicolumn{1}{r}{(0.1)} \\
\multicolumn{1}{l}{Missing} &
  \multicolumn{1}{|r}{-3.5***} &
  \multicolumn{1}{r}{-3.5***} &
  \multicolumn{1}{r}{-3.5***} &
  \multicolumn{1}{r}{-3.5***} &
  \multicolumn{1}{r}{-3.4***} \\
\multicolumn{1}{l}{} &
  \multicolumn{1}{|r}{(0.1)} &
  \multicolumn{1}{r}{(0.1)} &
  \multicolumn{1}{r}{(0.1)} &
  \multicolumn{1}{r}{(0.1)} &
  \multicolumn{1}{r}{(0.1)} \\
\multicolumn{1}{l}{Controls} &
  \multicolumn{1}{|r}{None} &
  \multicolumn{1}{r}{Linear} &
  \multicolumn{1}{r}{Bins} &
  \multicolumn{1}{r}{Max} &
  \multicolumn{1}{r}{Lender} \\
\multicolumn{1}{l}{Observations} &
  \multicolumn{1}{|r}{2,653,296} &
  \multicolumn{1}{r}{2,653,296} &
  \multicolumn{1}{r}{2,653,296} &
  \multicolumn{1}{r}{2,653,296} &
  \multicolumn{1}{r}{2,653,296} \\
\multicolumn{1}{l}{Adjusted R-squared} &
  \multicolumn{1}{|r}{0.02} &
  \multicolumn{1}{r}{0.02} &
  \multicolumn{1}{r}{0.02} &
  \multicolumn{1}{r}{0.18} &
  \multicolumn{1}{r}{0.35} \\
\cline{1-6}
\end{tabular}

\footnotesize{
Note: Each column reports results from an OLS regression of interest rate on race/ethnicity indicators and controls as indicated in the "Controls" row. Coefficients are reported in percentage points. The excluded race/ethnicity category is non-Hispanic White. Race/ethnicity categories are defined in Appendix Section \ref{sec:racecat}. Controls, when present, consist of credit score, DTI, and CLTV variables. "Linear" indicates these controls enter linearly; "Bins" indicates they enter as a series of indicators for each dimension (10 point ranges for credit score and 5 point ranges for DTI and CLTV); "Max" indicates the controls are all interactions of the indicators used in "Bins"; "Lender" indicates the controls are all interactions of the controls in "Bins," plus lender indicators. These regressions are run on standard purchase applications with zero discount points.
}
\end{table}

	\end{landscape}
}

\end{document}